\documentclass[review,12pt,3p,times]{elsarticle}

\usepackage{amssymb}

\usepackage{lineno}
\usepackage{setspace}
\usepackage[utf8]{inputenc}
\usepackage{amsmath}
\usepackage{overpic}

\usepackage{booktabs}
\usepackage{bm}
\usepackage{comment}
\usepackage{float}
\usepackage{gensymb}
\usepackage{graphicx}
\usepackage{multirow}
\usepackage{nomencl}
\usepackage{natbib}
\usepackage{tikz}
\usepackage{ulem}
\setcitestyle{square,comma,numbers,sort&compress}
\biboptions{numbers,sort&compress}

\usepackage[version=4]{mhchem}
\usepackage{siunitx}
\usepackage{subfigure}
\usepackage{subcaption}
\usepackage{textcomp}
\usepackage{longtable,tabularx}
\usepackage{ulem}
\usepackage{verbatim}
\usepackage{xcolor}
\usepackage{soul}
\usepackage{xpatch}

\makenomenclature
\usepackage{etoolbox}
\renewcommand\nomgroup[1]{%
  \item[\bfseries
  \ifstrequal{#1}{A}{}{%
  \ifstrequal{#1}{B}{Greek symbols}{%
  \ifstrequal{#1}{C}{Abbreviations}{}}}%
]}

\begin{document}

\begin{frontmatter}



\title{Numerical investigation of three-dimensional effects of cavitating flow  in a venturi-type hydrodynamic cavitation reactor}

\author[inst1]{Dhruv Apte*}
\author[inst2]{Mingming Ge}
\author[inst1,inst3]{Olivier Coutier-Delgosha}

\affiliation[inst1]{organization={Kevin T. Crofton Department of Aerospace and Ocean Engineering},
            addressline={Virginia Tech}, 
            city={Blacksburg},
            postcode={24060}, 
            state={VA},
            country={USA}}
\affiliation[inst2]{organization={
National Observation and Research Station of Coastal Ecological
Environments in Macao,Macao Environmental Research Institute, Faculty of Innovation Engineering, Macau
University of Science and Technology},
            city={Macau SAR},
            postcode={999078}, 
            country={China}}
\affiliation[inst3]{organization={Univ. Lille, CNRS, ONERA, Arts et Metiers ParisTech},
            addressline={Centrale Lille, FRE 2017 - LMFL - Laboratoire de Mecanique des fluides de Lille}, 
            city={Kampe de Feriet},
            postcode={F-59000}, 
            state={Lille},
            country={France}}   

\begin{abstract}
The concept of Hydrodynamic Cavitation (HC) has emerged as a promising method for wastewater treatment, bio-diesel production and multiple other environmental processes with Venturi-type cavitation reactors showing particular advantages. However, numerical simulations of a venturi-type reactor with an elucidated  explanation of the underlying flow physics remain inadequate. The present study numerically investigates and analyzes the flow inside a venturi-type reactor from both global cavity dynamics and localized turbulence statistics perspectives. Some models in the Detached Eddy Simulation (DES) family are employed to model the turbulence with the study initially comparing 2D simulations before extending the analysis to 3D simulations. The results show that while URANS models show significantly different dynamics as a result of grid refinement, the DES models show standard flow dynamics associated with cavitating flows. Nevertheless, significant discrepancies continue to exist when comparing the turbulence statistics on the local scale. As the discussion extends to 3D calculations, the DES models are able to well predict the turbulence phenomena at the local scale and reveal some new insights regarding the role of baroclinic torque into the cavitation-vortex interaction.The findings of this study thus contribute to the fundamental understandings of the venturi-type reactor.    
\end{abstract}




\end{frontmatter}


\section{Introduction} \label{introduction}
Cavitation is a multiphase flow phenomenon defined by the formation and subsequent collapse of vapor bubbles in a liquid caused by rapid changes in pressure. The sudden collapse of these vapor bubbles as they exit the low-pressure zone results in shock waves and erosion. These events are often detrimental for hydraulic applications like turbo-machinery and propeller blades as the shock and erosion result in noise and material damage, thus adversely impacting their efficiency. Conversely, the energy release as a result of the bubble collapse has also been harnessed for drilling jets for rock erosion in geothermal energy reservoirs \citep{hutli2017experimental}, process intensification in biodiesel production \citep{chuah2017kinetic, laosuttiwong2018performance} and for wastewater treatment \citep{wang2021hydrodynamic, saxena2018enhanced}. As HC is artificially generated in a hydrodynamic cavitation reactor (HCR), the design of a HCR to render it highly effective plays a pivotal role in HC technology. Numerous cavitating devices have been proposed as the geometry for HCR e.g. rotating-type \citep{sun2021multi, xia2024numerical}, vortex-type \citep{mittal2023intensifying, sarvothaman2024evaluating} and Venturi-type \citep{ge2022dynamic, mukherjee2022design}. Venturi-type HCRs are widely-used due to their relatively simple geometry. In a venturi-type HCR, HC occurs in the diverging part of the reactor in the direction of the flow.       

The occurrence of hydrodynamic cavitation is controlled by a non-dimensional parameter termed as cavitation number, $\sigma =  (p_{\infty}-p_{v})/(0.5\rho_{\infty} u_{\infty}^2 )$. Here, $p_{\infty},\rho_{\infty},u_{\infty}$ refer to the free-stream pressure, density and velocity respectively. In a classical view, cavitation can be separated into three fundamental regimes as the cavitation number ($\sigma$) decreases. The first stage is the cavitation inception, where low amounts of vapor are produced for a very short amount of time as the pressure of the working fluid drops below the vapor pressure. Since the bubble nuclei are extremely small to be visible, normally inception is detected using acoustic signals. The inception process continues as long as the liquid is able to sustain the tension of the microbubbles. As the $\sigma$ decreases, these bubbles grow visible and shed in a transition from sheet-like structures to intermittent cloud-like structures. It has been observed previously that the transition is triggered by formation of a re-entrant jet \citep{callenaere2001cavitation}. The shed cavity travels downstream whereupon entering the high-pressure region, collapses releasing shock waves. Further reduction in $\sigma$ results in a vapor cavity large enough to engulf the object travelling through the liquid. This phenomenon is termed as supercavitation and can also be generated by injecting non-condensable gas around the body. While there has been considerable experimental work conducted to investigate cavitating flows, the work can be augmented by conducting numerical simulations, as they provide access to visualizing the entire flow field in the machine.

Different numerical methodologies have been proposed to simulate cavitating flows. They can be roughly classified into Lagrangian and Eulerian approaches: In the Lagrangian approach, the field properties of the liquid phase are obtained using the Eulerian conservation equations, while the bubble dynamics are captured using the compressible Rayleigh-Plesset Equation. Lyu \textit{et al.} \citep{lyu2021modeling} used the Lagrangian approach to model the simulation of a cylindrical bubble cloud interacting with a pressure wave. Wang \textit{et al.} \citep{wang2022numerical} used a similar approach to assess the erosion risk in a cavitating axisymmetric nozzle. They concluded by stating that the erosion damage risk is amplified by the oscillation and collapse of near-wall bubbles generated near the cavity closure line.   While the Lagrangian approach can capture more flow details like discrete bubble clusters, it has two significant drawbacks. Firstly, the computational cost of such simulations is extremely high as the method tracks a large number of bubbles. Secondly, the approach assumes that cavitation is depicted solely by bubbles. Thus, the modelling strategy does not consider the cavity-liquid interface dynamics once the vapor cavity develops, a principal component for sheet and cloud cavitation.

The second approach, the Eulerian approach uses the Eulerian conservation equation to model both the liquid and vapor phases. Generally, the equations are coupled together with a void transport equation to account for cavitation. Several studies have previously proposed source terms for the void transport equation \citep{merkle1998computational, kunz2000preconditioned, schnerr2001physical}. These Transport-Equation Models (TEM) are coupled with a turbulence model to account for the interaction between cavitation and turbulence. Several studies have been conducted to investigate the cavitation-turbulence coupling using a diverse array of models. These models range from the Large Eddy Simulation (LES), where turbulent scales larger than a filter are resolved to the Reynolds-Averaged Navier-Stokes (RANS) models where all the turbulence scales are modelled.  Some studies \citep{gnanaskandan2016large} have used Large-Eddy Simulation (LES) to model the transition from sheet cavitation to cloud cavitation over a wedge. They observed multiple smaller vapor cavities shed apart from the primary cavity at the leading edge. They concluded by stating a strong co-relation between the adverse pressure gradient, the re-entrant jet and formation of cloud vapor cavity. Trummler \textit{et al.} \citep{trummler2020investigation} used implicit LES to investigate the cloud cavitation shedding and the re-entrant jet mechanism in a step nozzle. They observed that the shedding was initiated by condensation shocks at lower cavitating numbers whereas cavitation at higher cavitation numbers is initiated by the re-entrant jet. They confirmed that the re-entrant jet is formed as a result of the pressure peak induced by the cloud collapse after the cloud detaches from the leading edge. Coutier-Delgosha \textit{et al.} \citep{coutier2003evaluation} used three RANS models, the standard k-$\epsilon$ RNG, the modified k-$\epsilon$ RNG and the k-$\omega$ models to simulate cavitation in a venturi nozzle. The study concluded by stating none of these models were able to accurately reproduce the flow unsteadiness and periodicity of the cloud cavity as observed in experiments. To alleviate this issue, they introduced an eddy viscosity limiter termed as the Reboud correction to prevent the over-dampening of the eddy viscosity and reproduce the periodic shedding of the cloud cavity. More recently, Vaca-Revelo $\&$ Gnanaskandan \cite{vaca2023numerical} conducted similar URANS calculations to simulate the sheet to cloud transition over a wedge. They noted the presence of a re-entrant jet forming at the cavity closure that accelerated due to pressure waves formed downstream. In addition to URANS and LES models, there also exist hybrid RANS-LES models where the models behave as LES model away from the wall and as a RANS model towards the wall. These models thus are able to provide a much more accurate representation of the flow than RANS but with lesser computational time than LES and thus provide an excellent alternative to both RANS and LES models. Several such models have been proposed since the idea’s conception with the most notable one being the Detached Eddy Simulation (DES) \citep{spalart1997comments}. In this method, the model switches from a RANS model in the boundary layer to a LES model through a switch function dependent on the grid size. The switch function enables a sudden switch of the models and a major issue arises when the near-wall grid filter becomes smaller than the RANS length scale but not refined enough for LES. This would lead to LES resolved turbulence, termed as Modelled Stress Depletion (MSD). To ameliorate the issue, shielding functions were proposed in the subsequent studies \citep{spalart2006new} termed as the Delayed DES (DDES) model and Shur \textit{et al.} \citep{shur2008hybrid}, called the Improved DDES model (IDDES) model. Bensow \citep{bensow2011simulation} conducted flow simulations for cavitation around a twisted hydrofoil comparing the DDES model with implicit LES and k-$\epsilon$ model with the Reboud correction. The study found that while the DDES simulation was predicted a weaker re-entrant jet than LES, it did not over-predict the turbulent eddy viscosity unlike RANS. However, all these studies focused solely on the cavity dynamics and the pertaining flow physics at the global scale and did not investigate the formation of cloud cavity at the local scale. Sedlar \textit{et al.} \citep{sedlar2016numerical} conducted cavitating flow simulations over a hydrofoil using DES and LES models and compared them to experimental data. They noted that while LES indeed predicted more vortical structures, both models were able to predict the decrease of cavity length to zero and seemed to describe the cavity break-up. They also stated that the ability of both LES and DES models could be influenced by the computational grids. However, none of the above-mentioned studies evaluated the abilities of the turbulence models to simulate cavitating flows and investigated the turbulence statistics on a local scale. Indeed, table \ref{tab:lit_review} shows some recent studies conducted regarding HC in a venturi-type HCR. Most of these studies either discuss the global flow behaviour or investigate the influence of the geometrical parameters like inlet and outlet angle, flow velocity etc and not focus on the cavitation-turbulence coupling or the cavitation-vortex interactions. In addition, these studies employ the RANS modelling strategies that are not able to accurately model the cavitating flow. These shortcomings were also highlighted in Pipp \textit{et al.} \citep{pipp2021challenges}. Apte \textit{et al.} \citep{apte2023numerical} conducted a comprehensive analysis of several hybrid RANS-LES models, including DES and DDES to simulate cavitating flow in a venturi nozzle and compared the models with experimental data using local profile stations but these simulations were 2D and thus could not capture the cavitation-vortex interaction that can be visualized using 3D simulations. 
This study, therefore aims to evaluates the ability of the Detached Eddy Simulation (DES), Delayed DES, and the Improved DDES (IDDES) models to simulate cavitating flow in a venturi-type HCR using both 2D and 3D simulations from the cavitation-vortex interaction on the global scale to comparing the turbulence statistics using profile stations on the local stations. The simulations are compared with high-fidelity X-ray Particle-Image Velocimetry (PIV) data obtained from Ge \textit{et al.} \citep{ge2022dynamic} and provide directions to the development of industrial-scale HCRs for wastewater treatment and biodiesel production. 

\begin{table}
    \centering
    \begin{tabular}{lcccc} \toprule
        Year & Turbulence model & Content & Ref. \\ \midrule
        2019 & k-$\omega$ SST & Effect of geometrical parameters on cavitation  & \citep{simpson2019modeling} \\
        2020 & Realizable k-$\epsilon$ & Global cavitating flow characteristics & \citep{bimestre2020theoretical} \\
        2020 & quasi-Direct Numerical Simulation (DNS) & Global cavitating flow characteristics & \citep{fang2020numerical} \\
        2021 & k-$\omega$ SST with density correction \citep{reboud1998two, coutier2003evaluation} & Global cavitating flow characteristics  & \citep{pipp2021challenges} \\
        2021 & k-$\omega$ SST  & Effect of geometrical parameters on cavitation & \citep{dutta2021novel} \\
        2022 & RNG k-$\epsilon$ with density correction \citep{reboud1998two, coutier2003evaluation} & Global cavitating flow characteristics   & \citep{malekshah2022dissolved} \\
        2024 & Realizable k-$\epsilon$ & Effect of geometrical parameters on cavitation  & \citep{liu2024design} \\ 
        2024 & k-$\omega$ SST & Effect of geometrical parameters on cavitation  & \citep{guo2024numerical} \\ \bottomrule
    \end{tabular}
    \caption{Studies conducted to investigate the cavitating flow field in a venturi-type HCR}
    \label{tab:lit_review}
\end{table}
\section{Numerical Model}
\subsection{Basic governing equations}
This work uses the Transport-Equation Model (TEM) approach where the liquid and vapor are strongly coupled, governed by the momentum and mass transfer equations defined as:
\begin{equation}
    \frac{\partial (\rho_{m} u_{i})}{\partial t} + \frac{\partial (\rho_{m} u_{i} u_{j})}{\partial x_{j}} = -\frac{\partial p}{\partial x_{j}} + \frac{\partial }{\partial x_{j}}[(\mu_{t} + \mu_{m})(\frac{\partial u_{i}}{\partial x_{j}} + \frac{\partial u_{j}}{\partial x_{i}} - \frac{2}{3} \frac{\partial u_{k}}{\partial x_{k}}\delta_{ij})]
\end{equation}
\begin{equation}
    \frac{\partial \rho_{l} \alpha_{l}}{\partial t} + \frac{(\partial \rho_{l}\alpha_{l} u_{j})}{\partial x_{j}} = \Dot{m}^{+} + \Dot{m}^{-}
\end{equation}
\begin{equation}
    \rho_{m} = \rho_{l} \alpha_{l} + \rho_{v} \alpha_{v} 
\end{equation}
\begin{equation}
    \mu_{m} = \mu_{l}\alpha_{l} + \mu_{v} \alpha_{v}
\end{equation}
Where $u_{j}$ is the velocity component in the $\textit{j}$th direction, $\rho_{m}$ and $\mu_{m}$ are the density and viscosity of the mixture phase respectively, $u$ is the velocity, $p$ is the pressure, $\rho_{l}$ and $\rho_{v}$ are the liquid and vapor density respectively, $\mu_{l}$ and $\mu_{v}$ are the liquid and vapor dynamic viscosity respectively while $\mu_{t}$ represents the turbulent viscosity. The terms $\dot{m}^{+}$ and $\dot{m}^{-}$ denote the source and sink terms respectively or the vapor destruction (condensation) and vapor formation (evaporation) terms respectively. The Merkle transport-equation model \citep{merkle1998computational} is used to define the source and sink terms. Other models can be utilized here, but in order to reduce the uncertainties associated with simulating cavitating flows, the cavitation model has been solely fixed to the Merkle model. The model allows treating the subprocesses separately through the different source and sink terms and, unlike other TEMs, is not derived from the Rayleigh-Plesset equation. Thus, the model averts the assumptions associated with the Rayleigh-Plesset equation. The terms are defined as:

\begin{equation}
    \Dot{m}^{+} = \frac{C_{prod} max(p-p_{sat},0) (1-\gamma)}{0.5 U_{\infty}^{2} t_{\infty}}
\end{equation}

\begin{equation}
    \Dot{m}^{-} = \frac{C_{dest} min(p-p_{sat},0) \gamma \rho_{l}}{0.5 U_{\infty}^{2} t_{\infty} \rho_{v}}
\end{equation}
Where $\gamma$ is the liquid volume fraction, \textit{p} and $p_{sat}$ are the pressure and saturation pressure respectively while $t_{\infty}$ and $U_{\infty}$ are the free stream time scale and free stream velocity respectively. The empirical constants  $C_{des}$ and  $C_{prod}$ are set as 1e-3 and 80 respectively. 
\subsection{Turbulence models}

The models in the Detached Eddy Simulation (DES) family are utilized in this study. The DES models were originally based on the Spalart-Allmaras model \citep{spalart1992one} where the wall distance $d_{w}$ is replaced by $\Bar{d}$ involving the grid-size $\Delta$:
\begin{equation}
    \Bar{d} = min(d_{w}, C_{DES} \Delta)
\end{equation}
Where $\Delta = \max(\Delta_{1}, \Delta_{2}, \Delta_{3})$. $C_{DES}$ is a coefficient calibrated in the decaying homogeneous turbulence. In the near-wall region, as $\Delta = \Delta_{1}$ and $\bar{d} \approx \bar{d}_{w}$, the model reduces to the RANS model while far away from the wall, $d_{w} \gg \Delta$, the model acts as a subgrid scale model with $\bar{d} = C_{DES} \Delta$. Later, Spalart \textit{et al.} \citep{spalart2006new} observed that the original DES model exhibited incorrect behavior in thick boundary layers as the grid spacing parallel to the wall becomes less than the boundary layer thickness $\delta$. This resulted in the DES switching to LES despite the resolved Reynolds stress not being fully developed. They proposed a corrected version of the DES length-scale as:

  \begin{equation}
      \Bar{d} = d_{w} - f_{d} max(0, d_{w} - C{DES} \Delta)
  \end{equation})
Where $f_{d}$ is a shielding function that takes the value of unity in LES-modelled zone and reduces to zero everywhere else. This model, termed as Delayed DES (DDES) was further improved by Shur \textit{et al.} \citep{shur2008hybrid} who combined it with wall modelling in LES (WMLES) to resolve the mismatch between inner modeled log layer and outer resolved log layer. The new model, termed as Improved DDES (IDDES) formulated the blending function such that:
\begin{equation}
    \Bar{d} = \Bar{f_{d}}(1+f_{e})d_{w} + (1-\Bar{f_{d}})C_{DES} \psi\Delta
\end{equation}
Where $\Bar{f_{d}}, f_{e}, \psi$ are empirical functions. In addition, Travin \textit{et al.} \citep{travin2002physical} proposed the DES formulation based on the k-$\omega$ Shear Stress Transport (SST) model. The full formulation was presented by Menter \textit{et al.} \citep{menter2003ten} where the shielding function $F_{DES}$ was proposed as:
\begin{equation}
    F_{DES} = max(1, \frac{k_{sgs}^{\frac{3}{2}}/ \epsilon_{sgs}}{C_{DES} \Delta}(1- F_{SST})
\end{equation}
Where $F_{SST}$ is a blending function from the SST model. Similar relations result with the DDES and IDDES models.  

\section{Test case}
\subsection{Simulation setup}
\begin{figure*}
\includegraphics[width=18cm, height=3cm]{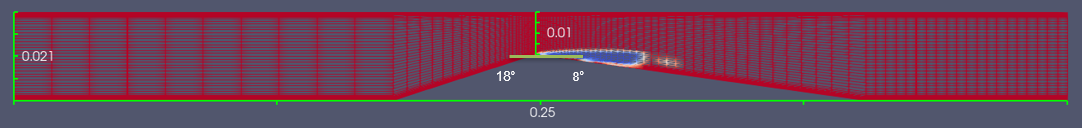}
\caption{The venturi-type geometry. All dimensions in mm. Here the blue region denotes the vapor bubbles region and the red region denotes water. Flow direction from left to right }
\label{fig:geometry}
\end{figure*}
The computational geometry is a converging-diverging (a \textit{venturi}) HCR from the X-ray high-speed Particle-Image Velocimetry (PIV) experiments of Ge \textit{et al.} \citep{ge2022dynamic} and numerical simulations of Apte \textit{et al.} \citep{apte2023numerical}. Venturi nozzles have been widely used to investigate cavitating flows for several applications like waste-water treatment \citep{agarkoti2023pilot}, bio-diesel production \citep{thakkar2022multi} and the hydrocarbon industry \citep{dehkordi2017cfd} and thus can be regarded as a standard geometry to model cavitating flows. Shown in Fig. \ref{fig:geometry}, it has an 18-degree convergent and 8-degree divergent angle. The sharp angle sat the throat enable the flow to be turbulent. The height of the venturi is 21 mm but it reduces to 11 mm at the throat. The inlet velocity is set to 8.38 m/s while the pressure is adjusted to meet the mean cavity length as seen in the experiments. To measure the extent of cavitation, a dimensionless number, the cavitation number ($\sigma$) is introduced as the ratio of pressure force to the inertial force. The cavitation number is defined as:
 \begin{equation}
     \sigma = \frac{2 (p-p_{v})}{\rho_{l} u^2}
 \end{equation}
Where $p_{v}$ is the vapor pressure,  $\rho_{l}$ is density of liquid, u is the section velocity and p is the outlet pressure.  Thus, lower the outlet pressure, the probability of cavitation and the resulting vapor cavity length will increase. In order to obtain the same cavity dynamics observed in experiments, the pressure at outlet is adjusted for every calculation to match the mean cavity length in the experiments. This ensures a consistency in comparison between numerical calculations and experiments and aids in focusing the study’s aspect solely on the turbulence modelling. The Re number is $1.8 \times 10^5$. Uniform velocity is implemented at the inlet.

\subsection{Numerical methods}
The numerical simulations are contained using OpenFOAM \citep{weller1998tensorial}, an open-sourced platform consisting of a multitude of solvers designed for specific fluid dynamics applications. The solver in this study is \textit{interPhaseChangeFoam}, an unsteady, multiphase and isothermal solver. Initially, a fully turbulent, non cavitating flow regime is run for 0.03s where the vaporization constant is set to zero. The regime is followed by a sinusoidal ramp for another 0.03s. During this ramp, there is no cavitation at the start while at the end of the ramp, there is full cavitation. Following this, a fully cavitating regime is launched for 1s. The focus of our results is on the final 1s of fully cavitating flow. The timestep used throughout this study is 1e-5. The equations are solved with the PIMPLE algorithm, a combination of the standard SIMPLE and PISO algorithms. 
Three 2-dimensional and three 3-dimensional grids are designed for the study (see Table \ref{tab:1}). To ensure the LES modelling region encompasses the cavity shedding and the near-wall regions are modelled by RANS, the meshes are designed in such a way that they are significantly refined in the cavitation region compared to the inlet or outlet regions. In addition, sample URANS (k-$\omega$ SST) calculations will also be conducted to evaluate the efficacy of the DES models. To re-emphasize, the objective of the calculations is to obtain the same mean cavity length that was observed in experiments (See Table \ref{tab:2})

\begin{table}[htbp]
    \centering
    \begin{tabular}{lccc} \toprule
        Name & Grid-size & Average $y^+$ \\
        \midrule
        35dot5k & 355 $\times$ 100 & 0.2250 \\
        51dot6k & 516 $\times$ 100 & 0.2168\\
        84k & 840 $\times$ 100 & 0.1794\\
        3dot3m & 330 $\times$ 100 $\times$ 100 & 0.218\\
        6m & 640 $\times$ 100 $\times$ 100 & 0.2884\\
        8dot4m & 700 $\times$ 120 $\times$ 100 & 0.1794\\ \bottomrule
    \end{tabular}
    \caption{Results of the grid-independence study for venturi-type reactor}
    \label{tab:1}
\end{table}

\begin{table}
    \centering
    \begin{tabular}{lcccc} \toprule
        Case & Cavitation Number ($\sigma$) & Mean cavity length & Shedding frequency\\ \midrule
         Experimental & 1.15 & 27 mm & 161 Hz\\ \bottomrule
    \end{tabular}
    \caption{Experimental data for comparison with Table \ref{tab:1}}
    \label{tab:2}
\end{table}
\section{Results}
The models are evaluated on their ability to evaluate periodic cavity shedding with the shedding zones distinctly in the LES modelling zones. The analysis is conducted using the DESModelRegions capability in OpenFOAM, that computes the field with the value ranging from 0 to 1 where 0 indicates a region modelled by the baseline RANS model and 1 where the region is modelled by LES. 
\begin{figure}[htp]

\centering
\includegraphics[width=.3\textwidth]{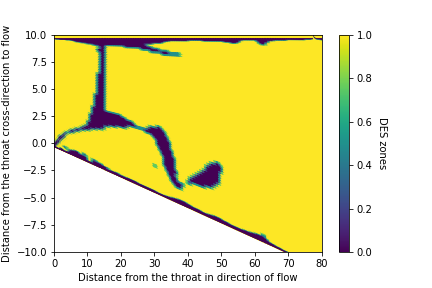}\hfill
\includegraphics[width=.3\textwidth]{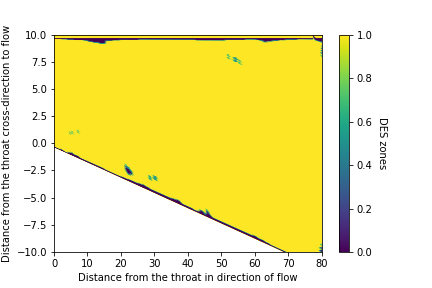}\hfill
\includegraphics[width=.3\textwidth]{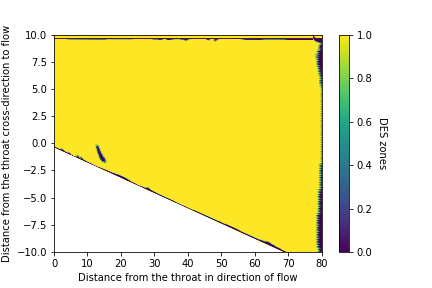}

\caption{LES and RANS zones in the diverging section of the venturi on the three two-dimensional meshes a) 35dot5k b) 51dot6k and c) 84k meshes. Here, 0 indicates region modelled by RANS (k-$\omega$ SST) and 1 indicates the region is modelled by LES}
\label{fig:zones}

\end{figure}
Fig \ref{fig:zones} shows the diverging section of the venturi-type reactor, where cavitation is expected to occur with the designation of the LES and RANS zones. All three grids demonstrate that almost the entirety of the section is modelled by the LES model with the exception of near the wall where, a thin but significant region is modelled by RANS. The meshes have been designed to ensure the wall are modelled coarsely as compared to the region away from the wall to reduce computing time.

\subsection{Evolution of cloud cavitation}
\begin{figure*}
\includegraphics[width=17cm, height=4cm]{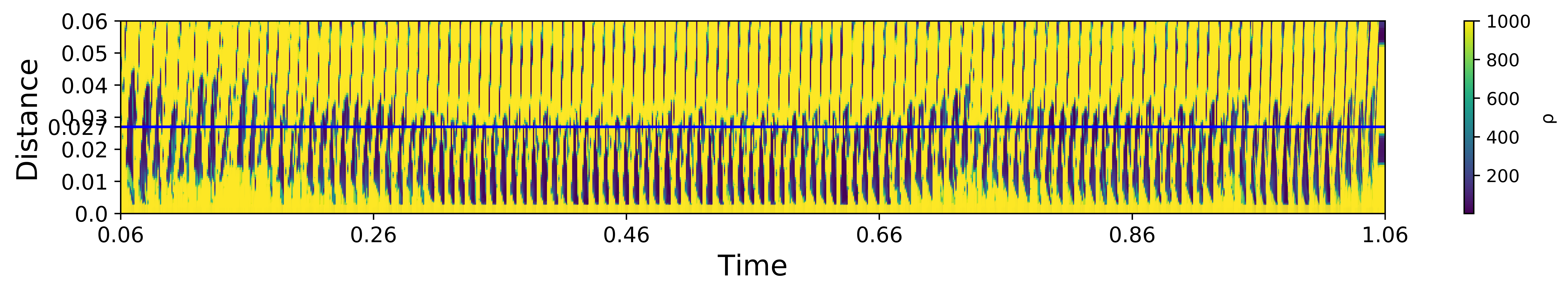}
\caption{Cavity Evolution Plot for the DES model on the 35dot5k mesh. The plot represents the minimum density value in each cross-section of the diverging part of the venturi over time on the x-axis and over its distance from the venturi throat on the y-axis. }
\label{fig:DEs35dot5kCV}
\end{figure*}

\begin{figure}[htp]
\centering
\includegraphics[width=.5\textwidth,  height=6cm]{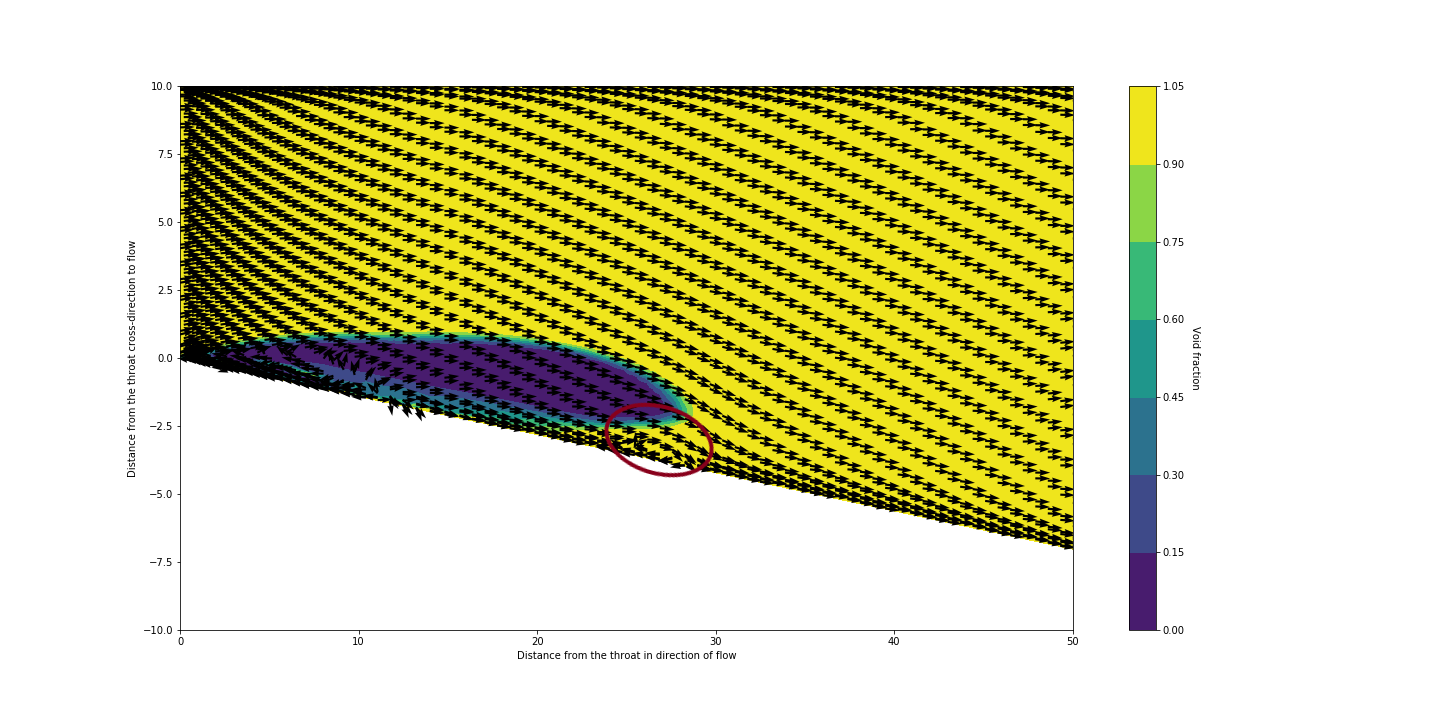}\hfill
\includegraphics[width=.5\textwidth,  height=6cm]{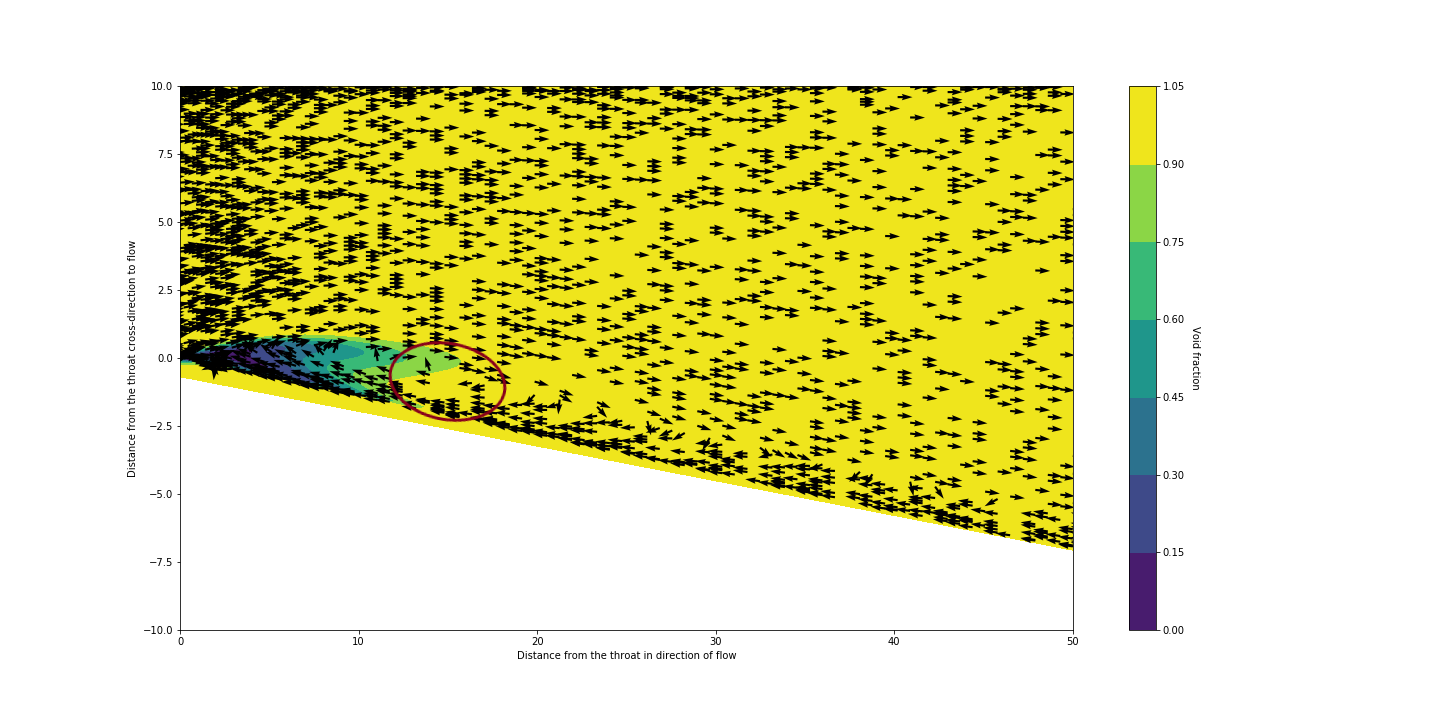}\hfill

\caption{Velocity streamlines (black arrows) with the arrowheads indicating direction of velocity vector super-imposed on mean void fraction plot for a) k- $\omega$ SST model and b) DES model simulations. The brown circled areas indicate the re-entrant jet as determined by the opposite velocity direction, much further downstream in the DES simulation than the URANS simulation. The cavity shapes also vary significantly.  }
\label{fig:streamlines}
\end{figure}

\begin{figure}[htp]

\centering
\includegraphics[width=.3\textwidth,  height=5cm]{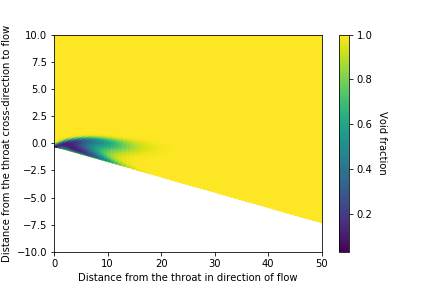}\hfill
\includegraphics[width=.3\textwidth,  height=5cm]{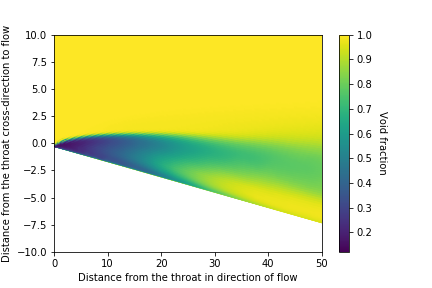}\hfill
\includegraphics[width=.3\textwidth,  height=5cm]{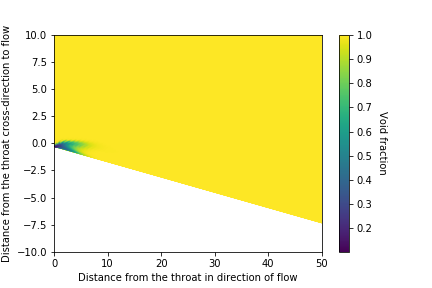}

\caption{Mean cavity shape for URANS simulations on the a) 35dot5k, b) 51dot6k and c) 84k cell meshes. In these figures, the yellow areas correspond to void fraction of 1 or pure water and a void fraction of zero corresponding to pure vapor. }
\label{fig:SST_cavity_shapes}

\end{figure}

In this section, the cloud cavities generated by these turbulence modelling calculations are investigated. This analysis is conducted by using the cavity evolution plots: these plots are plotted on a distance from the throat vs time axes and are color-coded by the minimum density in each cross-section of the diverging part of the venturi-type HCR. Studying the cavity evolution plot provides a glimpse into the primary concepts of cloud cavitation.  Fig \ref{fig:DEs35dot5kCV} demonstrates the cavity evolution plot of DES model on the 35dot5k mesh. 
\begin{figure*}
\includegraphics[width=17cm, height=8cm]{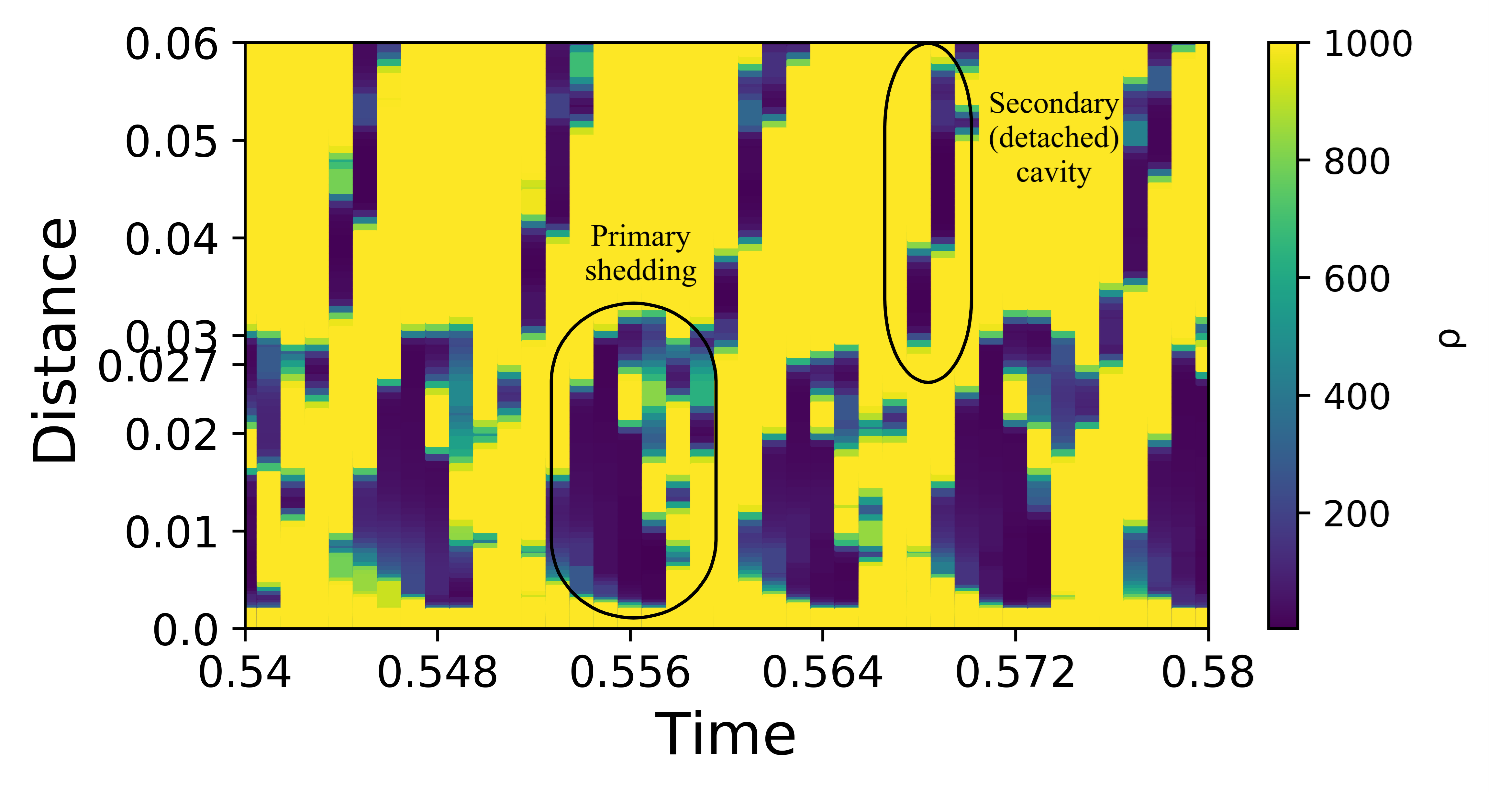}
\caption{Zoomed-in Cavity Evolution plot demonstrating the development of primary cavity near the throat and a secondary detached cavity further downstream. The primary cavities have a near-identical length, approximately 27 mm.}
\label{fig:DEs35dot5kCV_zoomed}
\end{figure*}
Fig \ref{fig:DEs35dot5kCV_zoomed} shows a zoomed-in portion of Fig \ref{fig:DEs35dot5kCV}. Here, the yellow regions represent water while the blue regions denote vapor. The distance is calculated as the distance between the point with minimum density and the throat. At periodic intervals near the throat, constant vapor clouds are observed. These clouds are followed by the advent of the re-entrant jet that breaks the primary cloud and results in formation of a thinner but longer secondary cavity downstream or the thin filaments observed in the figure. Fig \ref{fig:DEs35dot5kCV} shows the secondary detached cavity is approximately 33 mm in length. Zooming in again to Fig \ref{fig:DEs35dot5kCV_zoomed} shows the region near the throat is also interspersed with water. This indicates the primary cavity collapses, forming a zone of water before inception restarts and the periodic cycle continues. 
\begin{figure*}
\includegraphics[width=17cm, height=7cm]{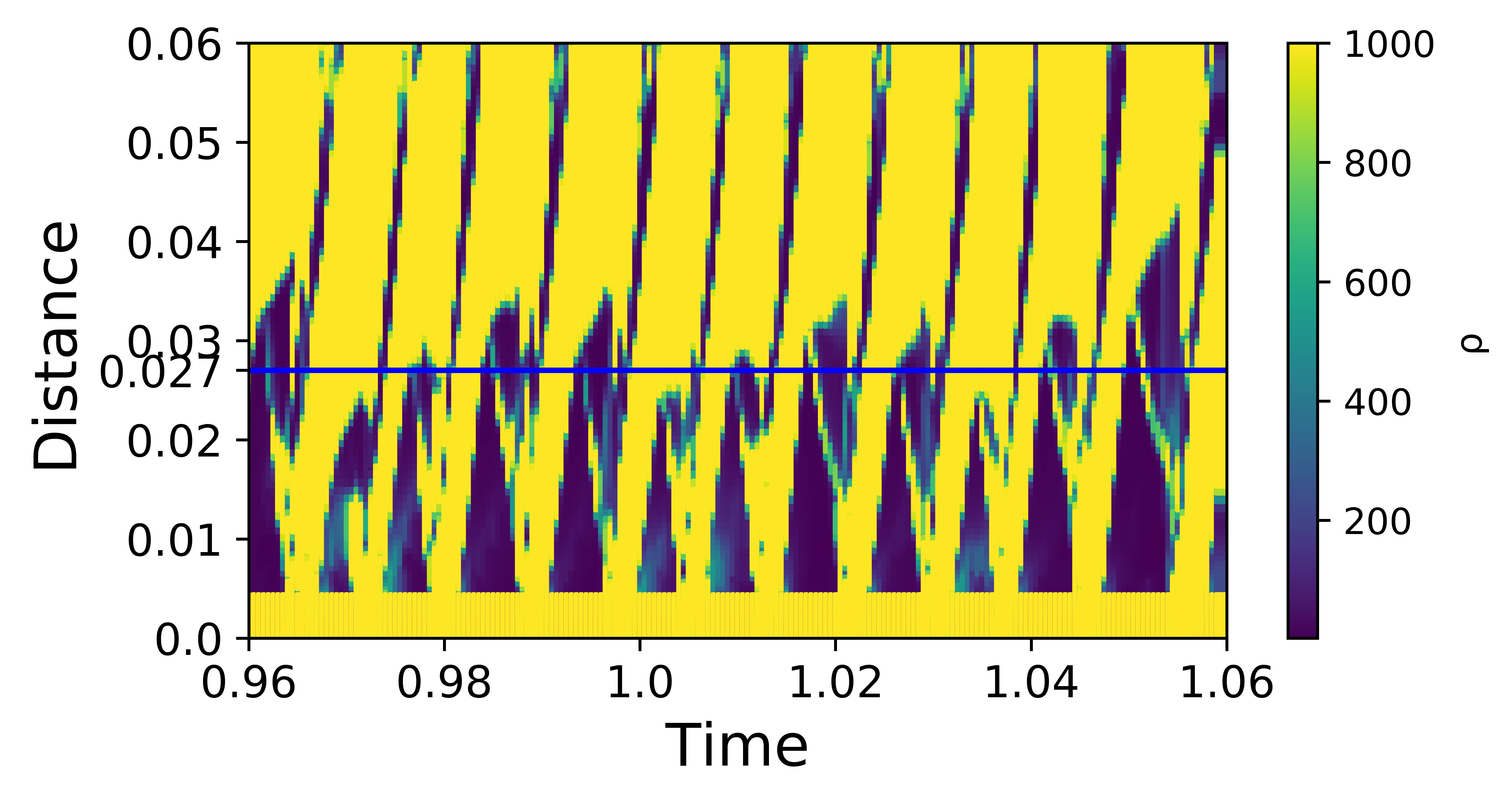}
\caption{Cavity Evolution Plot for the Delayed DES (DDES) model on the 35dot5k mesh captured for the last 0.01s. The plot is very similar to the DES model plot (Fig \ref{fig:DEs35dot5kCV}) but the cavities here are not uniformly 27 mm in length. }
\label{fig:DDES35dot5k}
\end{figure*}
\begin{figure*}
\includegraphics[width=17cm, height=6cm]{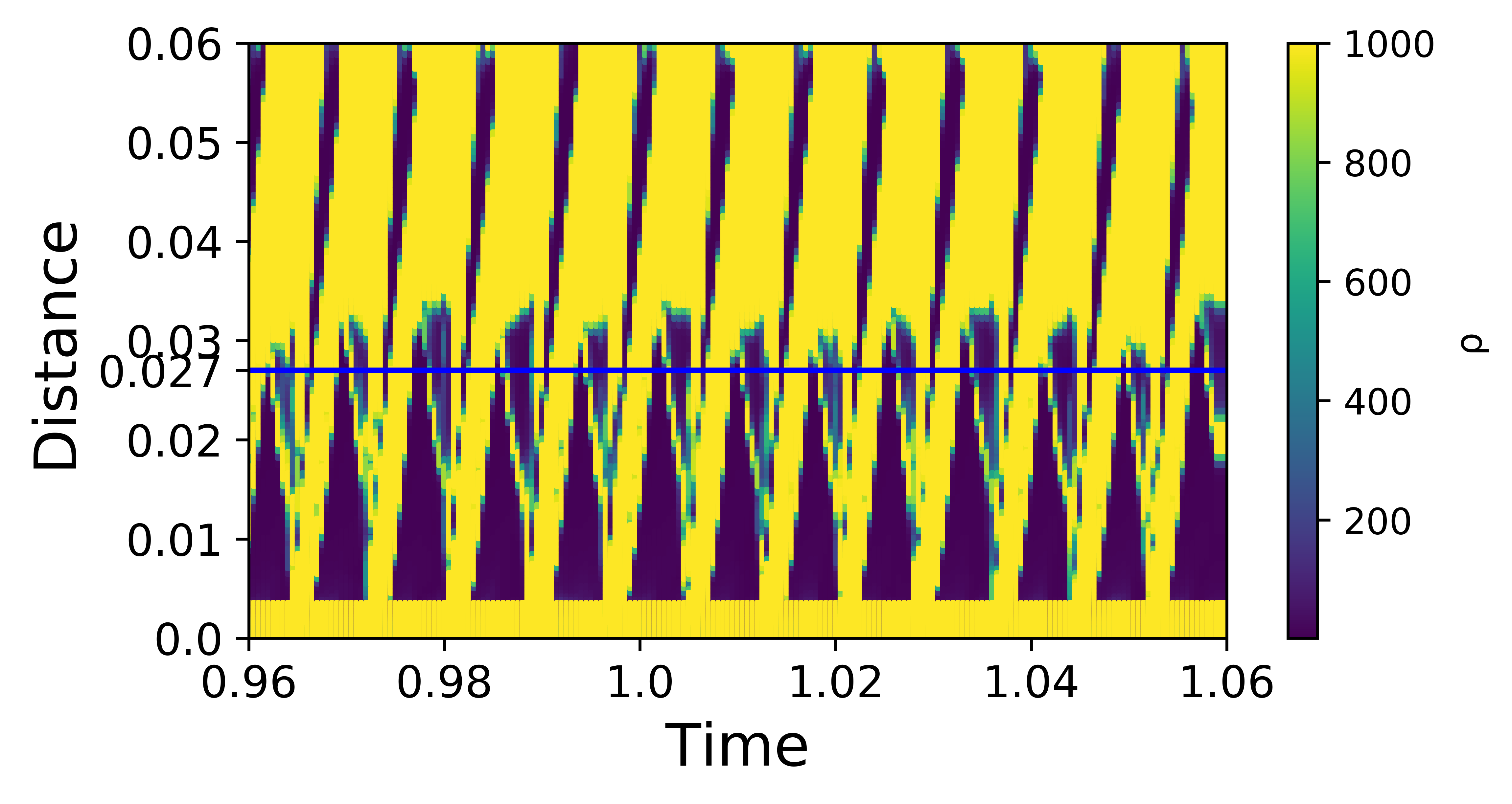}
\caption{Cavity Evolution Plot of Improved DDES model on the 35dot5k mesh. Here, unlike the DDES model, the cloud cavities are of uniform length. }
\label{fig:IDDES35dot5k}
\end{figure*}
Fig \ref{fig:DDES35dot5k} shows the cavity evolution plot for the Delayed DES (DDES) model, expected to improve the modelling in the transition zone. Here, the cavity dynamics shown are similar to the ones exhibited in the DES model (Fig \ref{fig:DEs35dot5kCV}) – the primary cavity at the throat followed by a re-entrant jet coming upstream, detaching the cavity, the detached cavity moving downstream and collapsing in a periodic manner. It is interesting to note that although the mean cavity length is 27 mm, not all cavities are equal in length. This is distinct from the Improved DDES (IDDES) model in Fig \ref{fig:IDDES35dot5k}, where all the cavities have identical lengths. 
\begin{figure*}
\includegraphics[width=17cm, height=4cm]{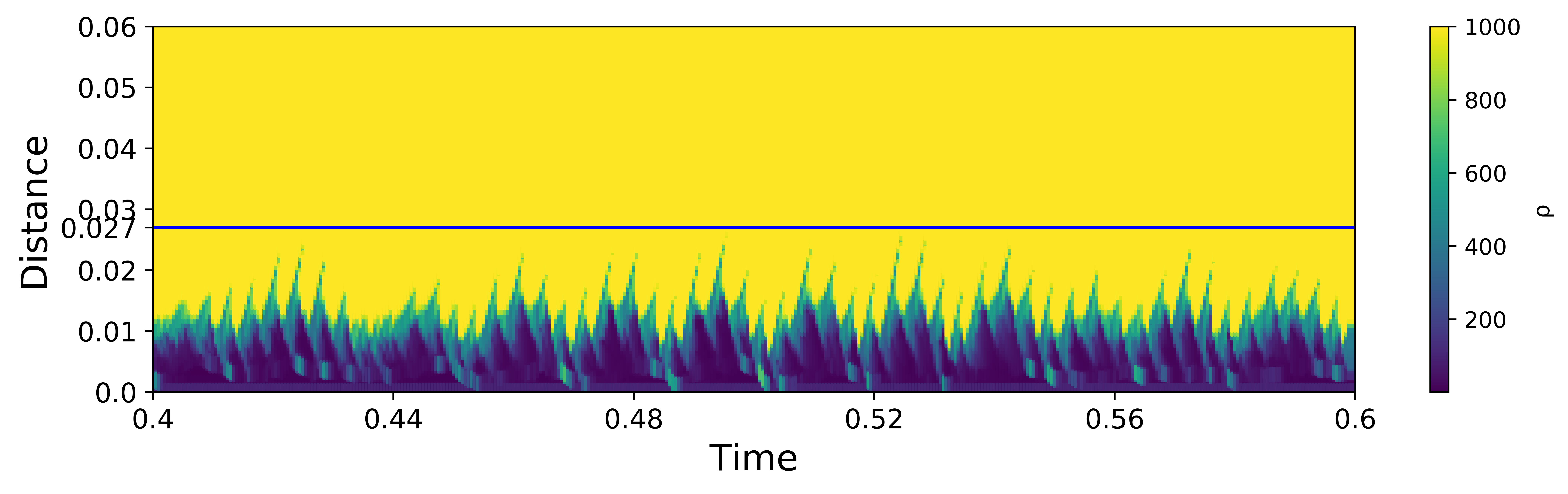}
\caption{Cavity Evolution Plot of k-$\omega$ SST model on the 35dot5k mesh. In sharp contrast from the DES model family plots, no periodic shedding and cavity detachment is observed. A singular cavity at the throat is observed across the simulation sample time, prompting further analysis.}
\label{fig:SST35dot5k}
\end{figure*}
To compare the overall performance of these models, the cavity evolution plot of the k-$\omega$ SST model calculation on the same mesh is also plotted in Fig \ref{fig:SST35dot5k}. Here, the observed cavity behavior is markedly different from the DES models. At the bottom of the plot, near the throat of the venturi-type HCR, a constant cavity is observed that does not break and re-form periodically and appears throughout the sample time as rather a singular cavity with its end forming ‘peaks’ periodically. This primary cavity is 0.01 mm long on average. Away from the throat, no secondary detached cavity or any vapor cavity is seen.
\begin{figure*}
\includegraphics[width=\textwidth, height=8cm]{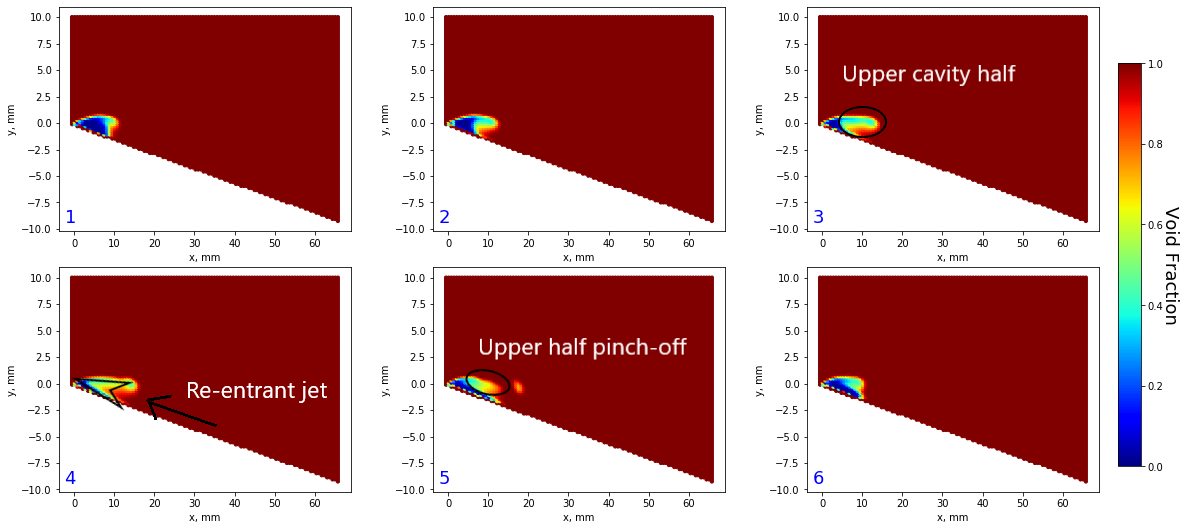}
\caption{Sequence of cavity growth and pinch off process for the k-$\omega$ SST simulation with void fraction field. For void fraction, 0 means pure vapor and 1 means pure water.}
\label{fig:cavity_grid}
\end{figure*}
A closer investigation on the sub-processes is shown by six successive snapshots in Fig \ref{fig:cavity_grid}. Fig \ref{fig:cavity_grid} (1) shows the primary cavity at the throat followed by its growth in (2). In (3), the upper half of the cavity is observed to be growing considerably faster than lower half until (4) where the cavity roughly resembles the shape of an arrowhead. (5) follows-up with the upper cavity half pinched off into a smaller detached cavity. This detached cavity is comparatively closer to the throat than the one observed in the DES model family simulations. (6) shows the cavity returning to its original shape with the detached cavity collapsed. The existence of cavity detachment from solely the upper cavity half and the formation of an arrowhead cavity shape shows that the re-entrant jet rushing upstream flows upwards as compared to the one observed in DES model family simulations. Fig \ref{fig:streamlines} provides a comparison of the re-entrant jet between DES and URANS simulations by plotting the time-averaged void fraction and then super-imposing the time-averaged velocity vectors on it. The mean cavity shapes are distinct with the URANS simulation having an arrowhead-shaped mean cavity (Fig \ref{fig:streamlines} (a)) while the DES simulation has a cloud shaped cavity (Fig \ref{fig:streamlines} (b)) in consistency with the experiments. The position of the re-entrant jet can be determined by the velocity vectors rushing upstream, as opposed to the flow direction: for the DES simulation, it is approximately 30 mm downstream, close to the wall but for the URANS simulation, the jet is found much closer to the throat, 12 mm downstream. The relatively upward position of the re-entrant jet in the URANS simulation results in it impacting the cavity midway, thus giving the arrow-head shape. This is also substantiated in the cavity evolution plot (Fig \ref{fig:SST35dot5k}) where the cavity peaks are intermittently filled with water. In addition, the re-entrant jet results in shedding of only the upper half of the cavity into a smaller detached cavity that collapses almost immediately. The position of the re-entrant jet is significantly different from previous URANS simulations observed in Apte \textit{et al.} \citep{apte2023numerical} and could be accounted by a combination of the highly refined grid in the adjoining region of the venturi in this case and, over-prediction of turbulent eddy viscosity by the URANS model.  
The results are replicated for the URANS simulations across all three meshes and all three simulations demonstrate the same arrowhead cavity shape as observed in Fig \ref{fig:SST_cavity_shapes}. While the size of the cavity varies across the simulations, the shape of the cavity remains constant. The observations are in clear contrast to the cavity shape obtained in the DES model family simulations. All the DES model simulations exhibit periodic shedding characteristics as discussed previously. Although these models show periodic shedding of cavities on a global level, further investigation is needed to examine the cavity dynamics predicted by these models on a much-localized scale. 
\subsection{Cavitation-Turbulence interaction on the local scale}
\begin{figure*}
\includegraphics[width=18cm, height=4cm]{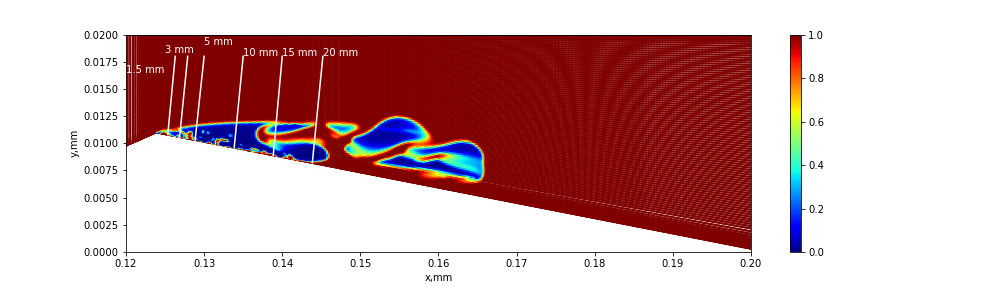}
\caption{Profiles used for local comparisons (dimensions in mm). These profiles are used to investigate whether a good agreement is obtained with experiments at a localized scale.}
\label{fig:profiles}
\end{figure*}

\begin{figure}
\centering
{\resizebox*{15cm}{!}{\includegraphics{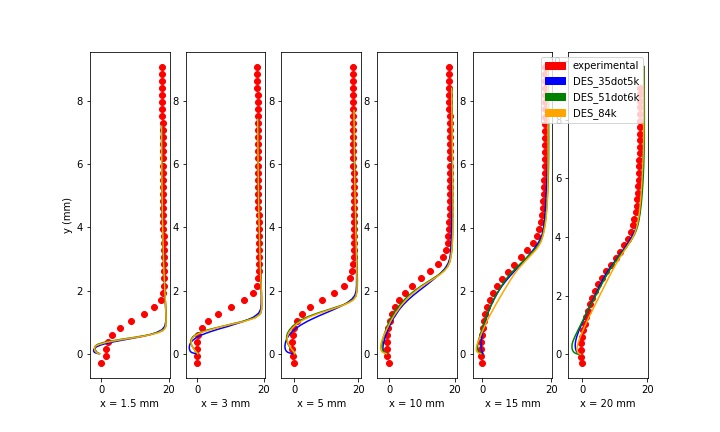}}}\hspace{8pt}

{\resizebox*{15cm}{!}{\includegraphics{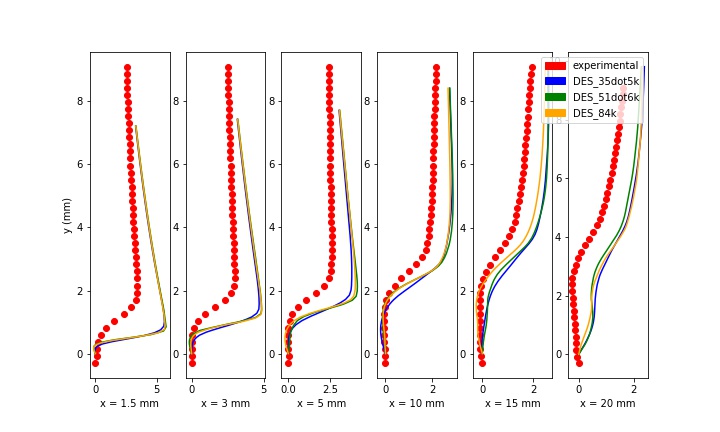}}}
\caption{Time-averaged velocity profiles in the (a) streamwise direction and (b) wall direction for the DES simulations. The red dots indicate the experimental data.} \label{fig:UMeanX_DES}
\end{figure}

To investigate the cavitation-turbulence interaction on the local scale, a local comparison is carried out at profiles 1.5 mm, 3mm, 5mm, 10mm, 15mm and 20 mm from the throat as shown in Fig \ref{fig:profiles}. The models are first evaluated across the three meshes separately. Fig \ref{fig:UMeanX_DES} shows the time-averaged velocity profiles in both the streamwise and wall direction for the DES model. While the profiles in streamwise direction have a good agreement with the experiments, the profiles in wall direction yield significant discrepancies downstream and away from the wall. Although the simulation on 84k cell mesh notably performs slightly better than the others, it still over-predicts the wall velocity away from the wall.   

\begin{figure}
\centering
{\resizebox*{15cm}{!}{\includegraphics{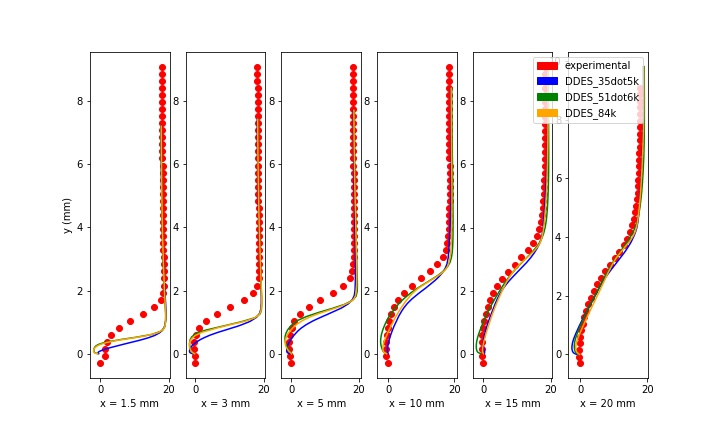}}}\hspace{8pt}

{\resizebox*{15cm}{!}{\includegraphics{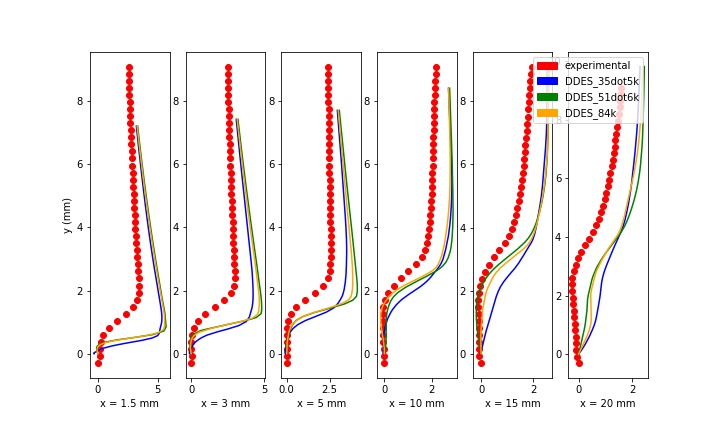}}}
\caption{Time-averaged velocity profiles in the (a) streamwise direction and (b) wall direction for the DDES simulations. The red dots indicate the experimental data.} \label{fig:UMeanX_DDES}
\end{figure}

\begin{figure}
\centering
{\resizebox*{15cm}{!}{\includegraphics{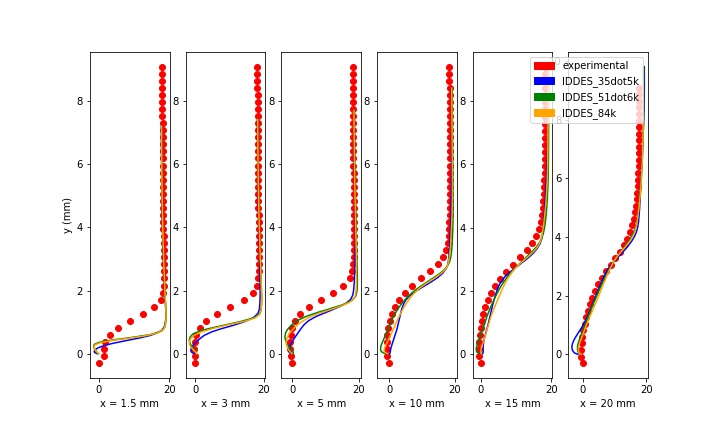}}}\hspace{8pt}

{\resizebox*{15cm}{!}{\includegraphics{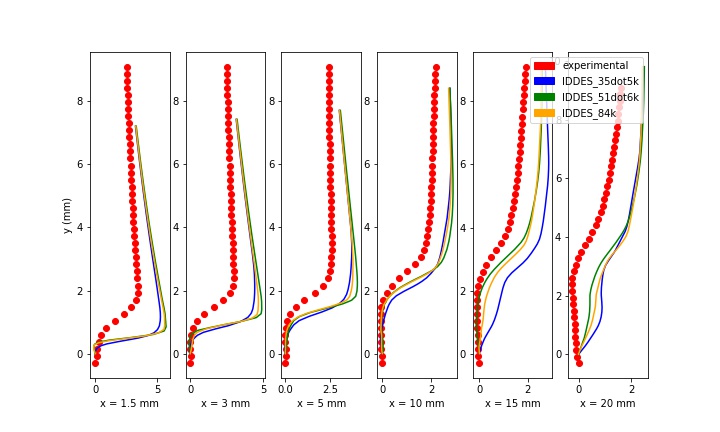}}}
\caption{Time-averaged velocity profiles in the (a) streamwise direction and (b) wall direction for the IDDES simulations. The red dots indicate the experimental data.} \label{fig:UMeanX_IDDES}
\end{figure}
\begin{figure}
\centering
{\resizebox*{15cm}{!}{\includegraphics{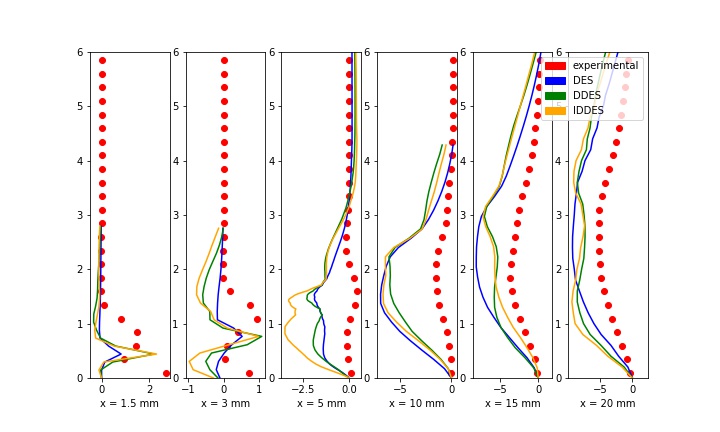}}}\hspace{8pt}

{\resizebox*{15cm}{!}{\includegraphics{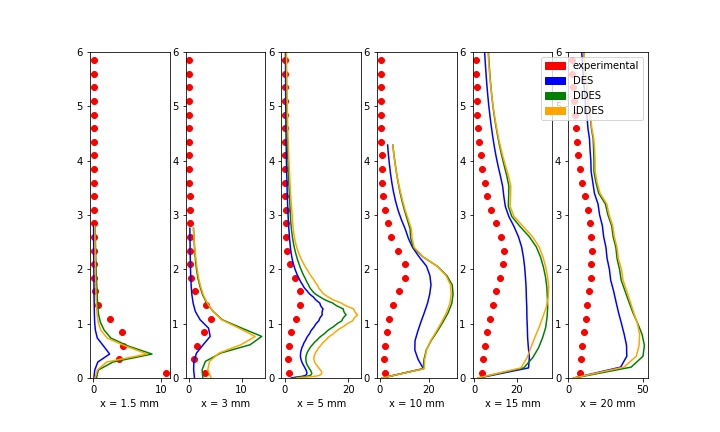}}}
\caption{Reynolds shear stress (a) and Turbulent Kinetic Energy (TKE) (b) profiles for all three models on the 84k cell mesh.} \label{fig:Tau12_84k}
\end{figure}

Improved models in the DES family like the DDES and IDDES models exhibit a similar trend as observed in Figs \ref{fig:UMeanX_DDES} and Fig \ref{fig:UMeanX_IDDES}. While the time-averaged velocities in streamwise direction agree well with the experimental data, the velocity profiles in the wall direction significantly over-predict the velocity. Near the throat, the models have slight differences, but these substantiate downstream and away from the wall. However, in both the DDES and IDDES, the 51dot6k and 84k cell mesh simulations align closer to the experiment than the 35dotk in the 15 mm and 20 mm profiles. 
\begin{figure}
\centering
{\resizebox*{15cm}{!}{\includegraphics{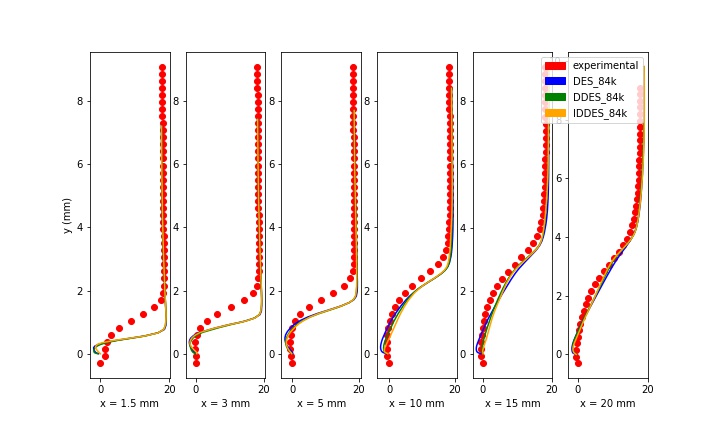}}}\hspace{8pt}

{\resizebox*{15cm}{!}{\includegraphics{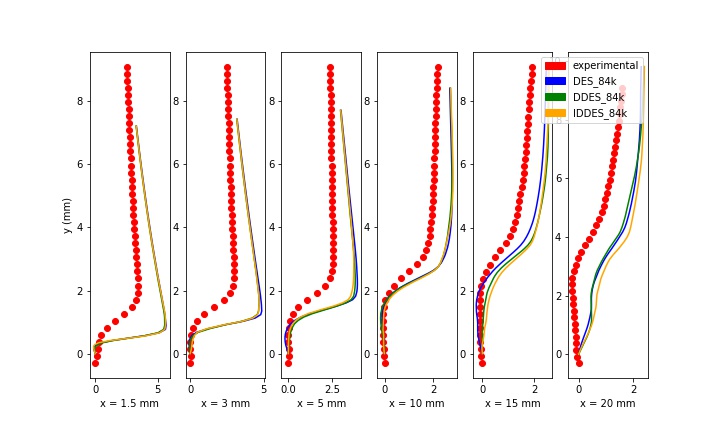}}}
\caption{Time-averaged velocity profiles in the (a) streamwise direction and (b) wall direction for all three models on the 84k cell mesh. Both direction profiles are identical for all three models.} \label{fig:UMeanX_84k}
\end{figure}
To evaluate the models reproducing the cavitation-turbulence interaction on a localized scale, the profiles of the three models are plotted in Fig \ref{fig:UMeanX_84k} on the 84k cell mesh. The models display identical behaviors in both the streamwise and wall direction profiles, including the discrepancies discussed previously regarding the wall direction profiles. 
The investigation on the local profiles is extended to the Reynolds shear stress and Turbulent Kinetic Energy (TKE). Indeed, the DES model family does not rely on the Boussinesq hypothesis here as the region of interest in these simulations is modelled by LES. Therefore, the data obtained is composed directly of the velocity fluctuations.
\begin{figure}
\centering
{\resizebox*{15cm}{!}{\includegraphics{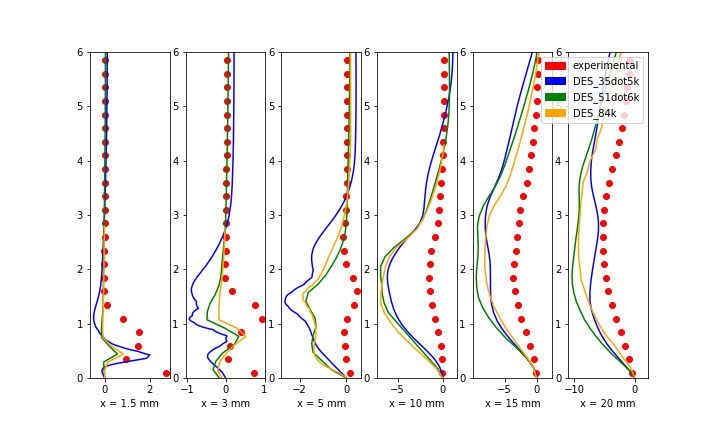}}}\hspace{8pt}

{\resizebox*{15cm}{!}{\includegraphics{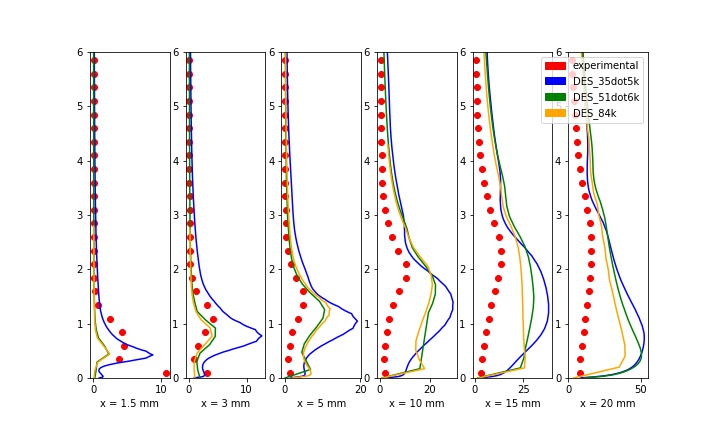}}}
\caption{Reynolds shear stress (a) and Turbulent Kinetic Energy (TKE) (b) profiles for the DES model on all three meshes. The red dots indicate experimental data.} \label{fig:Tau12_DES}
\end{figure}
Fig \ref{fig:Tau12_DES} presents the profile data of the Reynolds shear stress and TKE for the DES model on all three meshes. Looking at the Reynolds shear stress plots (Fig \ref{fig:Tau12_DES} (a)) the DES model underpredicts the shear stress near the throat but over-predicts the shear stress magnitudes in the downstream profiles. The TKE plots inform more data about the simulations and the underlying flow physics. The TKE in each of the subplots reaches its maximum value at the vapor-water interface forming a small ‘peak’. Tracing the peaks across the plots would reproduce the cloud cavity shape. Observing the simulation profiles shows that the DES models on the 51dot6k and 84k meshes have a better agreement with the experiments than the 35dot5k mesh, especially near the throat. These agreements deteriorate in the downstream profiles where all simulations display similar behaviors of having widespread discrepancies with the experimental data. Since the TKE and Reynolds shear stress is composed of the velocity fluctuations in both streamwise and wall directions, the discrepancies discussed previously are reflected again in the turbulence data plots. A further evaluation is conducted to analyze the three models on the fine 84k cell mesh in Fig \ref{fig:Tau12_84k}. While all three models are unable to reproduce the cavity flow dynamics downstream for both Reynolds shear stress and TKE, they show a similarity in the turbulent behavior upstream with some over-prediction. The TKE peaks for the models are closer to the wall as compared to the experimental data. Thus, the cavity width or the cavity-water interface, specifically for the numerical models is not the same as for the experiments. 

\subsection{3D effects on Cavitation-turbulence interaction}

\begin{figure}
\centering
{\resizebox*{15cm}{!}{\includegraphics{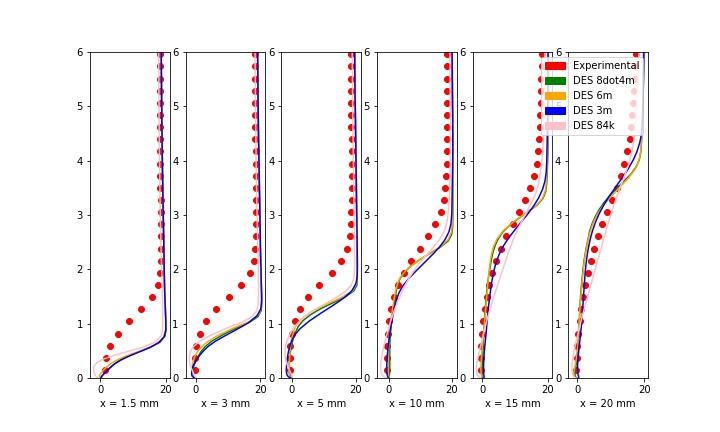}}}\hspace{8pt}
{\resizebox*{15cm}{!}{\includegraphics{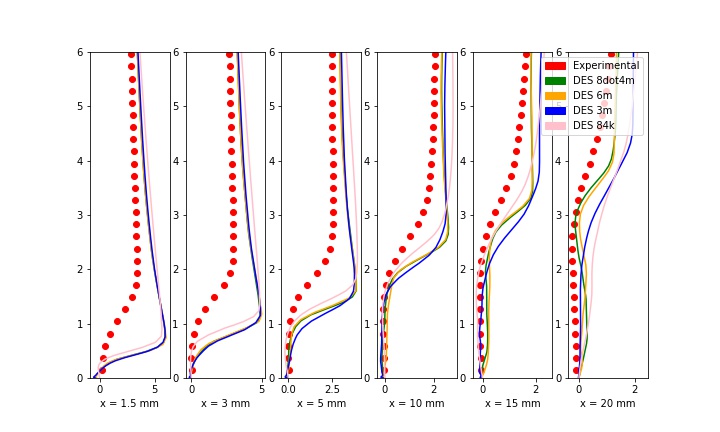}}}
\caption{Time-averaged velocities in a) streamwise direction and b) wall direction for 2D and 3D DES calculations. The pink lines correspond to the DES calculation on the 84k mesh,a 2D simulation while the other simulations correspond to the 3D simulations } 
\label{fig:U_2D_vs_3D}
\end{figure}

To investigate the impact of 3D effects on the cavitating flow, comparisons between 2D and 3D simulations have been conducted, specifically in for the time-averaged velocities, Reynolds shear stress and TKE. For the 3D cases, the data has been taken in the mid-plane where the cavity is expected to reach its maximum length. Fig \ref{fig:U_2D_vs_3D} depicts the time-averaged velocities for both streamwise and wall direction profiles for the 3D cases and the 2D simulation on the finest mesh. The velocity profiles for all cases in stream-wise direction are almost identical. Significant differences are observed, however in the wall direction velocity profiles as in Fig \ref{fig:U_2D_vs_3D} (b) which magnify downstream. At the x=20 mm profile, the DES simulation on 3 million cells mesh exhibits the the behaviour farthest from the experiments followed by the 2D simulation. It is also noted that the velocity profiles for DES simulations on both 6 million cells mesh and 8.4 million cells mesh are identical.    

\begin{figure}
\centering
{\resizebox*{15cm}{!}{\includegraphics{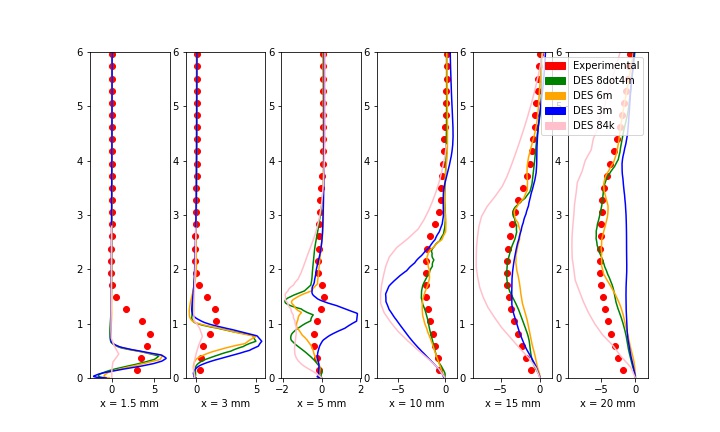}}}\hspace{8pt}
{\resizebox*{15cm}{!}{\includegraphics{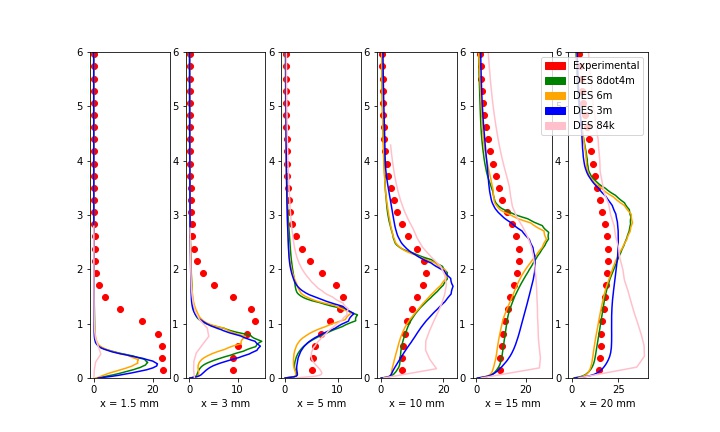}}}
\caption{Reynolds shear stress (a) and Turbulent Kinetic Energy (TKE) (b) for 2D and 3D DES calculations. The pink lines correspond to the DES calculation on the 84k mesh,a 2D simulation while the other simulations correspond to the 3D simulations } 
\label{fig:Turb_2D_vs_3D}
\end{figure}

The turbulence properties for both 2D and 3D simulations are presented in Fig \ref{fig:Turb_2D_vs_3D}. Fig \ref{fig:Turb_2D_vs_3D} (a) depicts the Reynolds shear stress profiles. Here, the 3D effects of turbulence can be noted distinctly. Near the throat, the 2D simulation under-predicts the stress significantly. A comparatively smaller difference appears between the 3D cases and the experimental data. Downstream, at the x =10 mm profile, both the 2D and the 3 million cell mesh 3D DES simulations appear to over-predict the Reynolds stress significantly while the other two 3D simulations are able to depict the Reynolds stress accurately. Further downstream, the 3 million cells mesh seem to under-predict the stress magnitude while the other two 3D calculations continue to predict the Reynolds stress accurately. It seems the 2D simulation is predicting a thinner cloud cavity than observed. This is substantiated in fig \ref{fig:Turb_2D_vs_3D} (b) which represents the Turbulent Kinetic Energy (TKE) profiles for the same simulations. Initially, near the throat, the 2D simulation under-predicts the magnitude of TKE while the 3D models are able to accurately predict the magnitude and the cavity behaviour. Downstream, the differences between 2D and 3D simulations become more pronounced. Here, while the 2D DES simulation not only over-predicts the TKE magnitude but also predicts a high value near the wall of the diverging section of the venturi rather than upwards, away from the wall. This indicates while the cavity length is equal, the cavity height is considerably different from the 3D simulations. It can be concluded that the cavity dynamics are significantly different between 2D and 3D simulations with the 3D simulations predicting the cavitation-turbulence interaction accurately.
 
\begin{figure}
\centering
{\resizebox*{15cm}{!}{\includegraphics{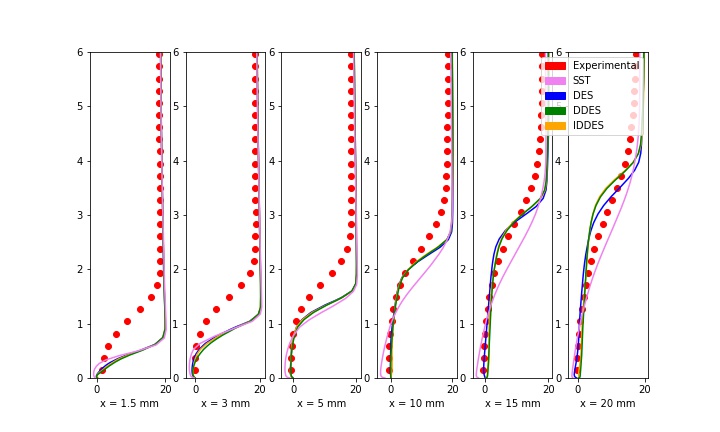}}}\hspace{8pt}
{\resizebox*{15cm}{!}{\includegraphics{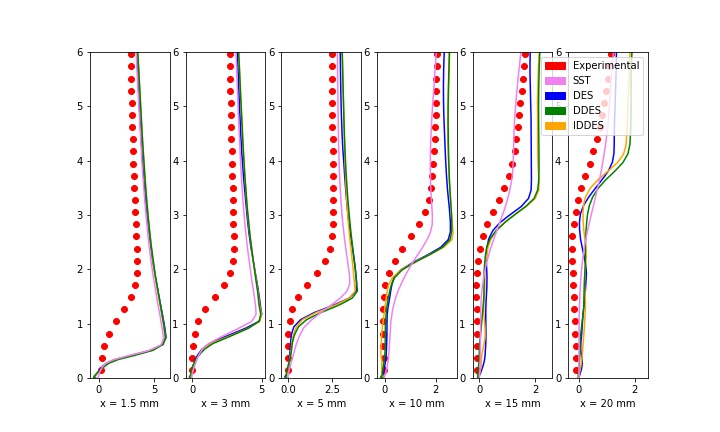}}}
\caption{Time-averaged velocity profiles in streamwise direction (a) and wall direction (b) for 3D SST and DES calculations. The models display identical behaviour, especially near the throat but display different profiles downstream} 
\label{fig:velocity_6m}
\end{figure}

Once a good agreement is noted between experiments and 3D simulations, the dynamics exhibited by different models on a three-dimensional mesh are discussed. Since both DES calculations on the 6 million cell mesh and 8.4 million cell meshes are identical as shown in Figs \ref{fig:U_2D_vs_3D} and \ref{fig:Turb_2D_vs_3D}, subsequent  simulations and analysis is conducted on the 6 million cells mesh. Fig \ref{fig:velocity_6m} depicts the time-averaged velocity profiles in both streamwise and wall directions. Near the throat, all models show identical behaviour with the sharp velocity jump closer to the wall section than experimental data in both directions. However, the SST simulation under-predicts the velocity downstream, especially in streamwise direction as observed in Fig \ref{fig:velocity_6m} (a). Similarly in Fig \ref{fig:velocity_6m} (b) the SST model exhibits different behaviour downstream as a steady increase in wall direction velocity rather than a small jump as observed in other cases. These differences will be compounded during the turbulence data profile analysis. 

\begin{figure}
\centering
{\resizebox*{15cm}{!}{\includegraphics{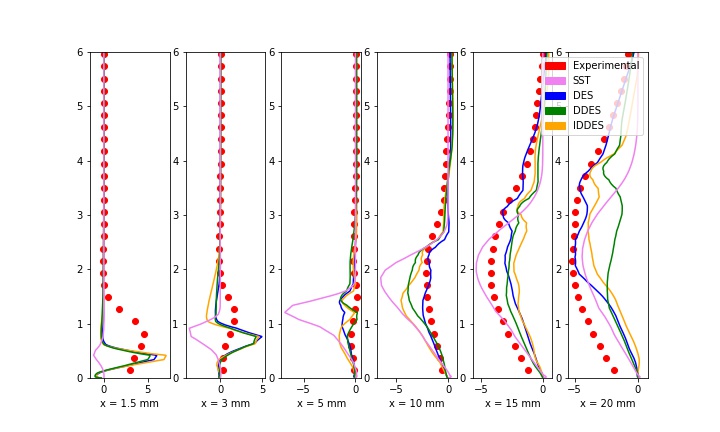}}}\hspace{8pt}
{\resizebox*{15cm}{!}{\includegraphics{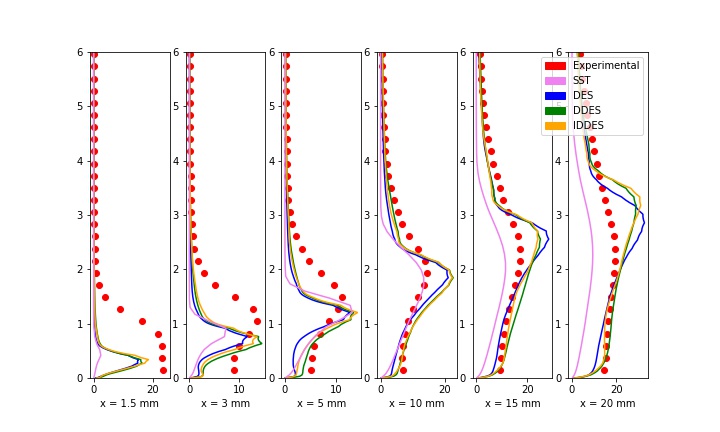}}}
\caption{Reynolds shear stress (a) and Turbulent Kinetic Energy (TKE) (b) for various 3D DES calculations and SST calculation on the 6 million cells mesh. The violet lines correspond to a SST simulation on the same mesh. While the SST over-predicts Reynolds stress and under-predicts the TKE, similar to the 2D calculations, slight differences are observed between the DES models on the 3D cases.} 
\label{fig:Turb_6m}
\end{figure}

\begin{figure}
\centering
{\resizebox*{15cm}{!}{\includegraphics{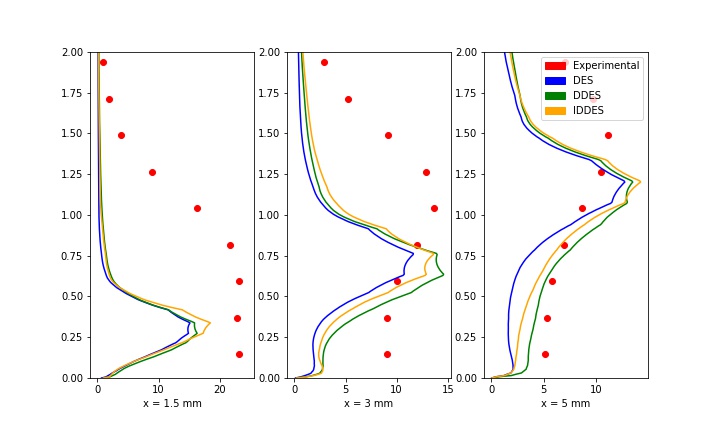}}}
\caption{Turbulent Kinetic Energy (TKE) profiles near the throat for various 3D DES calculations, zoomed in from Fig \ref{fig:Turb_6m} (b). The proposed improvements to the DES models in the form of the DDES and IDDES model are observed to affect the turbulence dynamics here, as the DES, DDES and the IDDES models show TKE values closer to the experiments in that order respectively.} 
\label{fig:Turb_6m_zoomed}
\end{figure}

Fig \ref{fig:Turb_6m} (a) presents the Reynolds stress profiles for the SST, DES, DDES and the IDDES models. The 3D SST simulation continues to exhibit the discrepancies observed in the 2D simulations. The DES simulations show identical behaviour near the throat but show slight differences downstream from the experimental data and each other. The experimental downstream profiles show where a single bump occurs, consisting of the minimum Reynolds stress value whereas the DES models show two bumps with the primary bumps in DES and IDDES models identical to the experimental one. Fig \ref{fig:Turb_6m} (b) shows the TKE profiles for the same simulations compared to experimental data. The proposed improvements to the DES models in the form of the DDES and IDDES models impacting the turbulent behaviour is observed near the throat. Here, the cavity is initiated and is very close to the wall. The proposed improvements in the DDES and IDDES models respectively are able to predict TKE values closer to the experimental plots as compared to the DES simulation, as observed in Fig \ref{fig:Turb_6m_zoomed}. On a general basis,it is observed that the DES models are able to well predict the TKE magnitude and the turbulent cavity behaviour here with slight differences as compared to the SST model. This agreement is in case due to the resulting modelling of the cavity region away from the wall by LES where the turbulent eddies are being resolved rather than modelled as is the case in SST.     

\subsection{Cavitation-Vortex interaction}
Previous studies have discussed the role of a unsteady periodic cavitating flow in the formation of vortex structures. For a better understanding of the cavitation phenomenon and its interaction with vortex formation, the vorticity transport equation is employed in the z-direction to investigate how its components are individually influenced:
\begin{equation}
    \frac{D\omega_{z}}{Dt} = [(\omega \cdot \nabla) V]_{z} - [\omega (\nabla \cdot V)]_{z} +  \left( \frac{\nabla \rho_{m} \times \nabla p}{\rho_{m}^{2}} \right)_{z} + [(\nu_{m} + \nu_{t}) \nabla^{2} \omega]_{z} 
\end{equation}\label{eq:VTE}

Here,the Left Hand Side (LHS) denotes the rate of vorticity change while the Right Hand Side (RHS) indicate the vortex stretching, vortex dilatation, baroclinic torque and viscous diffusion of vorticity terms respectively. The vortex stretching term describes the stretching and tilting of a vortex due to the velocity gradients. The vortex dilatation describes the expansion and contraction of a fluid element. The baroclinic torque is a result of the misalignment between the pressure and velocity gradients. The viscous diffusion term can be ignored in high Reynolds number flows \cite{dittakavi2010large} and thus is omitted in the study.

\begin{figure*}[htbp]
\includegraphics[width=\textwidth, height=8cm]{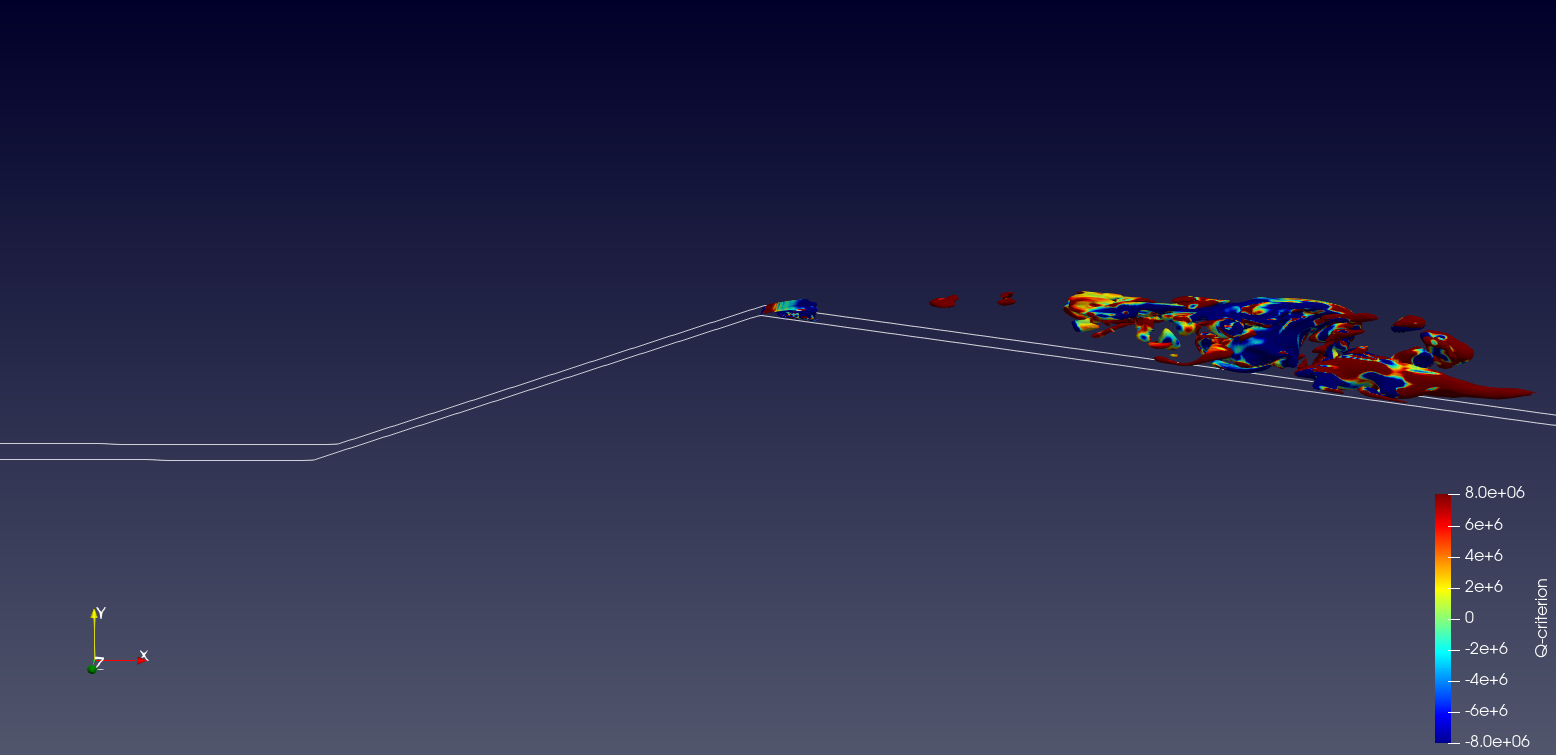}
\caption{Q-criterion contour for the cavity initiation stage}
\label{fig:ciq}
\end{figure*}

\begin{figure}[htbp]
     \centering
     \begin{subfigure}
         \centering
         \includegraphics[scale=0.35]{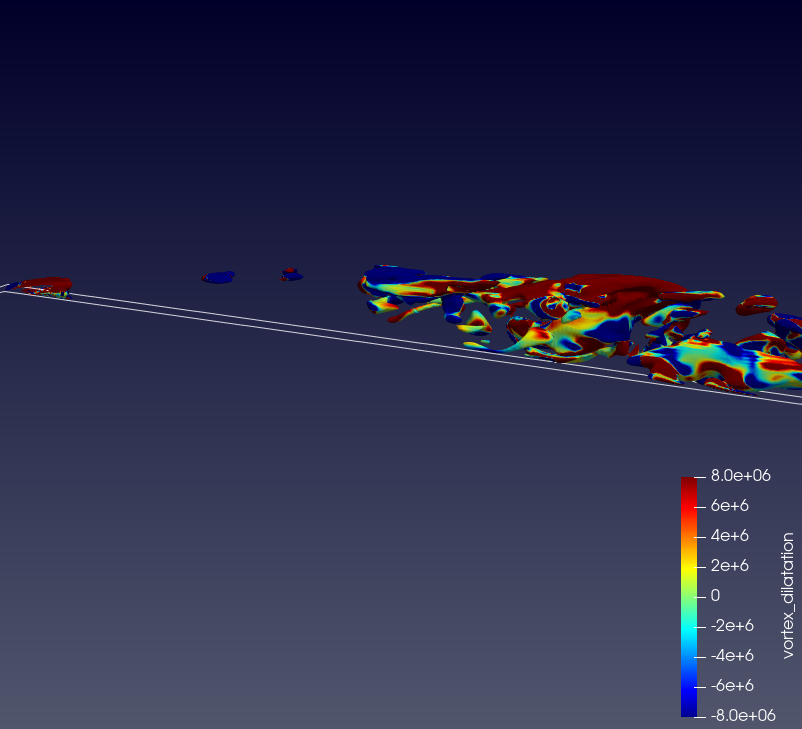}
         \label{fig:vd0}
     \end{subfigure}
     \hfill
     \begin{subfigure}
         \centering
         \includegraphics[scale=0.35]{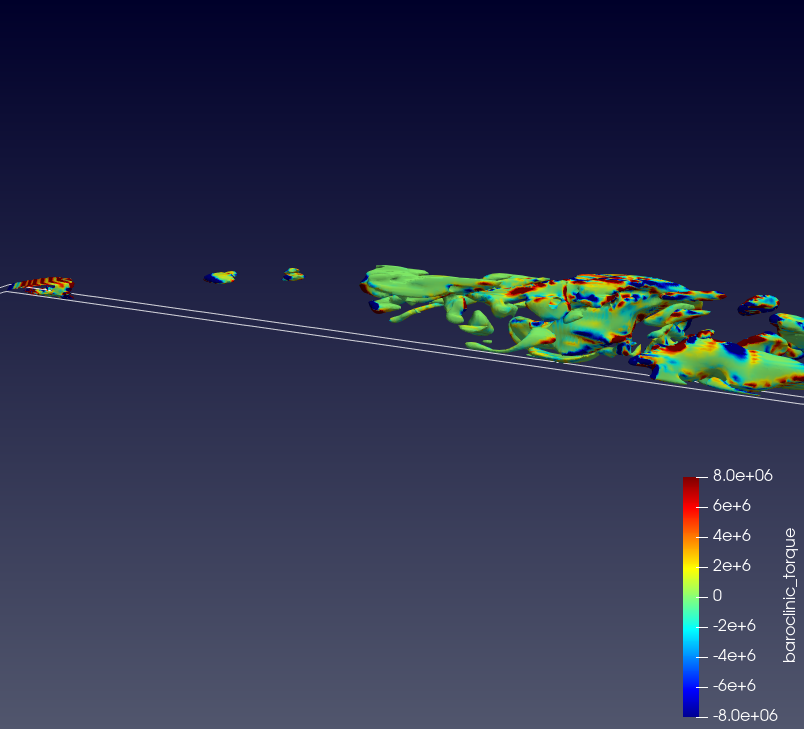}
         \label{fig:bt0}
     \end{subfigure}
        \caption{Starting top left, in clockwise direction,contours of void fraction (contour of $\alpha$ = 0.6, vortex stretching, baroclinic torque and vortex dilatation terms as a cavity is initiated as seen from the diverging-section of the venturi. Here, the white outline represents the bottom wall.}        
        \label{fig:ci2}
\end{figure}

\begin{figure*}[htbp]
\includegraphics[width=\textwidth, height=8cm]{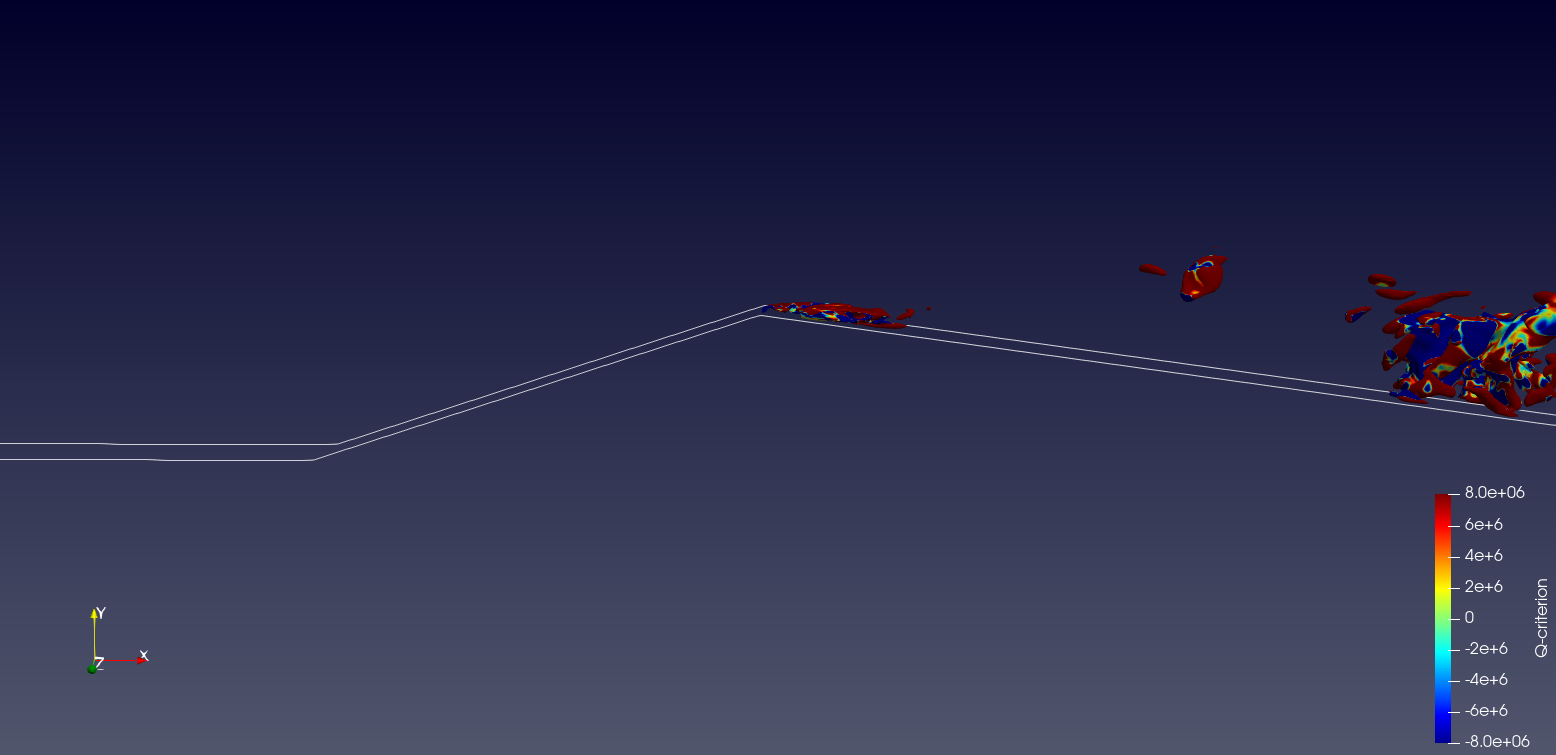}
\caption{Q-criterion contour as the cavity grows bigger}
\label{fig:ceq}
\end{figure*}

\begin{figure}[htbp]
     \centering
     \begin{subfigure}
         \centering
         \includegraphics[scale=0.35]{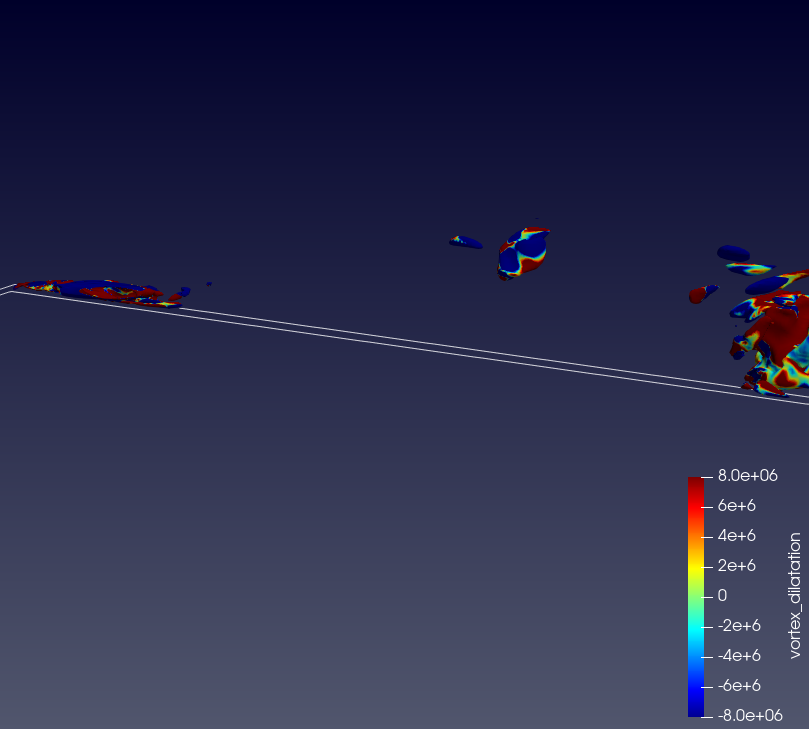}
         \label{fig:vd1}
     \end{subfigure}
     \hfill
     \begin{subfigure}
         \centering
         \includegraphics[scale=0.35]{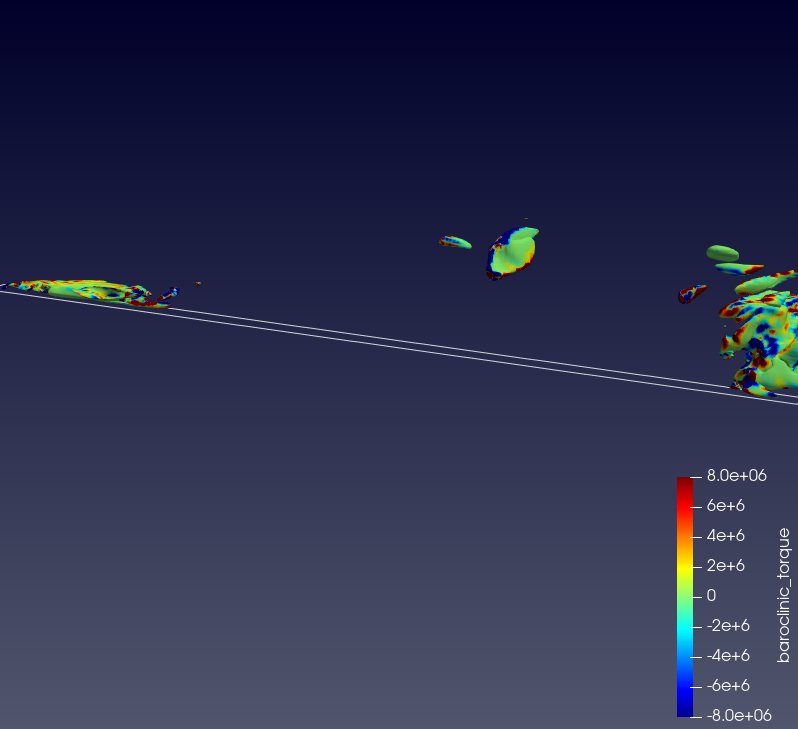}
         \label{fig:bt1}
     \end{subfigure}
        \caption{Starting top left, in clockwise direction,contours of void fraction (contour of $\alpha$ = 0.6), vortex stretching, baroclinic torque and vortex dilatation terms as the cavity grows bigger.}        
        \label{fig:ce2}
\end{figure}

The equation clearly shows the effects of velocity and pressure gradients created as a result of cavitating flow on the vorticity. The equation is further simplified as:
\begin{equation}
    \omega_{z} = \frac{\partial V_{y}}{\partial x} - \frac{\partial V_{x}}{\partial y}
\end{equation}
\begin{equation}
    [(\omega \cdot \nabla) V]_{z} = \omega_{x} \frac{\partial V_{z}}{\partial x} + \omega_{y} \frac{\partial V_{z}}{\partial y} + \omega_{z} \frac{\partial V_{z}}{\partial z}
\end{equation}
\begin{equation}
    [\omega (\nabla \cdot V)]_{z} = \omega_{z} (\frac{\partial V_{x}}{\partial x}+\frac{\partial V_{y}}{\partial y}+ \frac{\partial V_{z}}{\partial z})
\end{equation}
\begin{equation}
     \left( \frac{\nabla \rho_{m} \times \nabla p}{\rho_{m}^{2}} \right)_{z} = \frac{1}{\rho_{m}^{2}} (\frac{\partial \rho_{m}}{\partial x} \cdot \frac{\partial p}{\partial y} - \frac{\partial \rho_{m}}{\partial y} \cdot \frac{\partial p}{\partial x})
\end{equation}
\begin{equation}
    Q = \frac{1}{2} (||\Omega_{ij}||^2 - ||S_{ij}||^2)
    \label{eq:Q}
\end{equation}
The vortex-distribution terms and corresponding vapor structures are presented in the following figures along with Q-criterion, that defines regions where the magnitude of vorticity is greater than the magnitude of the rate of strain as stated in Eq \ref{eq:Q}. Here, a snapshot of the venturi from the front is taken of contours of the $\alpha$ = 0.6 followed by various components of the vorticity transport equation along with Q-criterion. The snapshots are taken from the IDDES calculation on the 6 million cells mesh which showed the most agreement with experimental data in the previous subsection.The snapshots are divided into six distinct stages of a periodic cavitating cycle. Figs \ref{fig:ciq} and \ref{fig:ci2} shows the initiation cycle. Here, an incipient cavity has started to manifest at the throat of the venturi and remnants of the cavity formed in the previous shedding cycle are observed. In fig \ref{fig:ciq}, the regions shaded by red indicate that the rate of rotation is greater than the strain rate, demonstrating the regions where vorticity is generated.  The vortex-stretching and vortex dilatation terms appear to dominate the vortex dynamics while the baroclinic torque is largely zero.Fig \ref{fig:ce2} shows the incipient cavity growing bigger as the pressure decreases. Fig \ref{fig:ceq} shows more vortices are being formed as the cavity grows larger.  The vortex shedding and dilatation terms continue to dominate the process with the baroclinic torque term continuing to barely influence the vortex dynamics process. As stated previously, the increase in velocity gradients and expansion of fluid element contribute to the larger roles played by the vortex stretching and vortex dilatation terms respectively.   

\begin{figure}[htbp]
     \centering
     \begin{subfigure}
         \centering
         \includegraphics[scale=0.35]{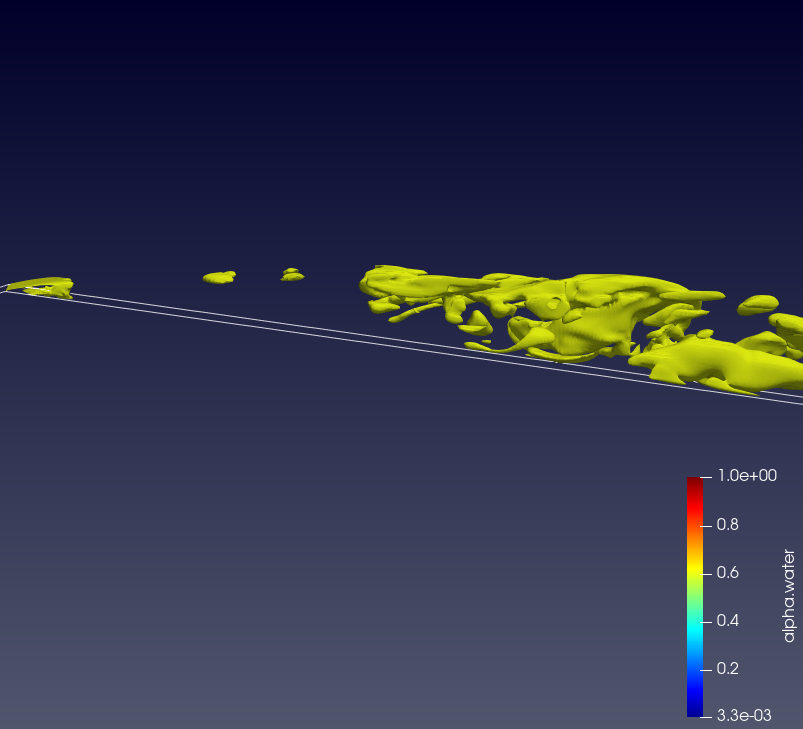}
         \label{fig:void_fraction0}
     \end{subfigure}
     \hfill
     \begin{subfigure}
         \centering
         \includegraphics[scale=0.35]{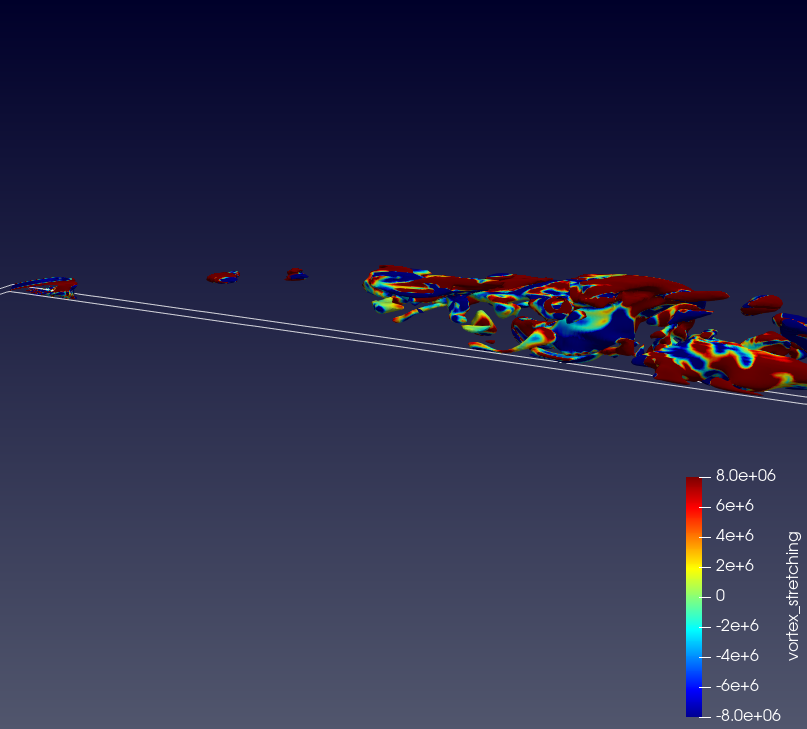}
         \label{fig:vs0}
     \end{subfigure}
        \label{fig:ci1}
\end{figure}

\begin{figure}[htbp]
     \centering
     \begin{subfigure}
         \centering
         \includegraphics[scale=0.35]{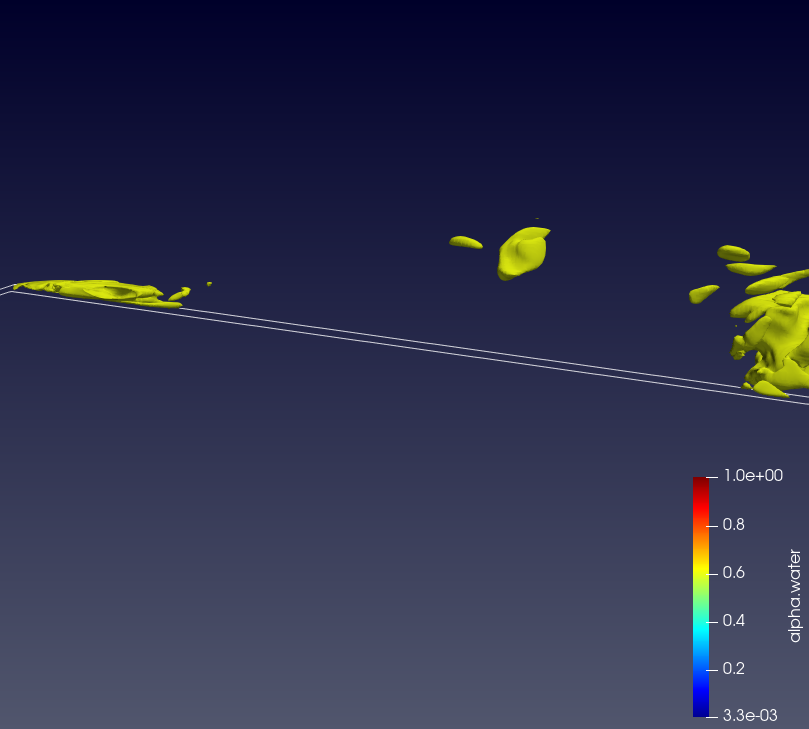}
         \label{fig:void_fraction1}
     \end{subfigure}
     \hfill
     \begin{subfigure}
         \centering
         \includegraphics[scale=0.35]{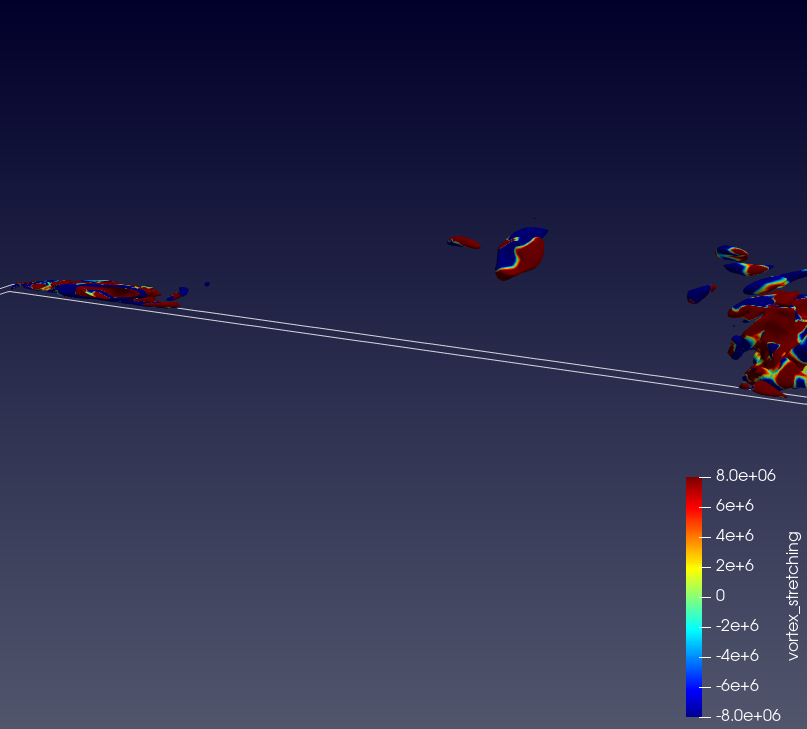}
         \label{fig:vs1}
     \end{subfigure}
       \label{fig:ce1}
\end{figure}
\begin{figure}[htbp]
     \centering
     \begin{subfigure}
         \centering
         \includegraphics[scale=0.35]{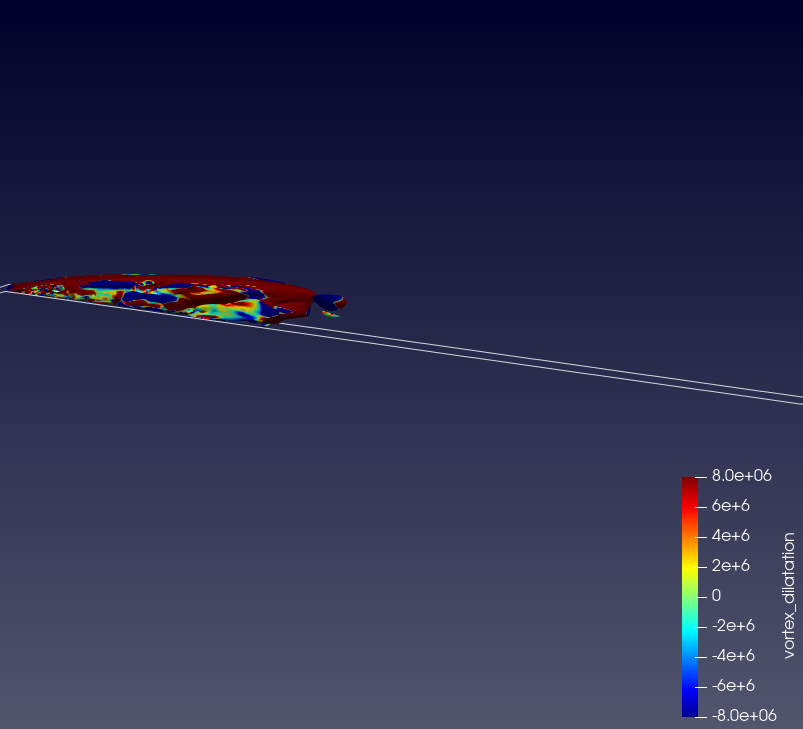}
         \label{fig:vd2}
     \end{subfigure}
     \hfill
     \begin{subfigure}
         \centering
         \includegraphics[scale=0.35]{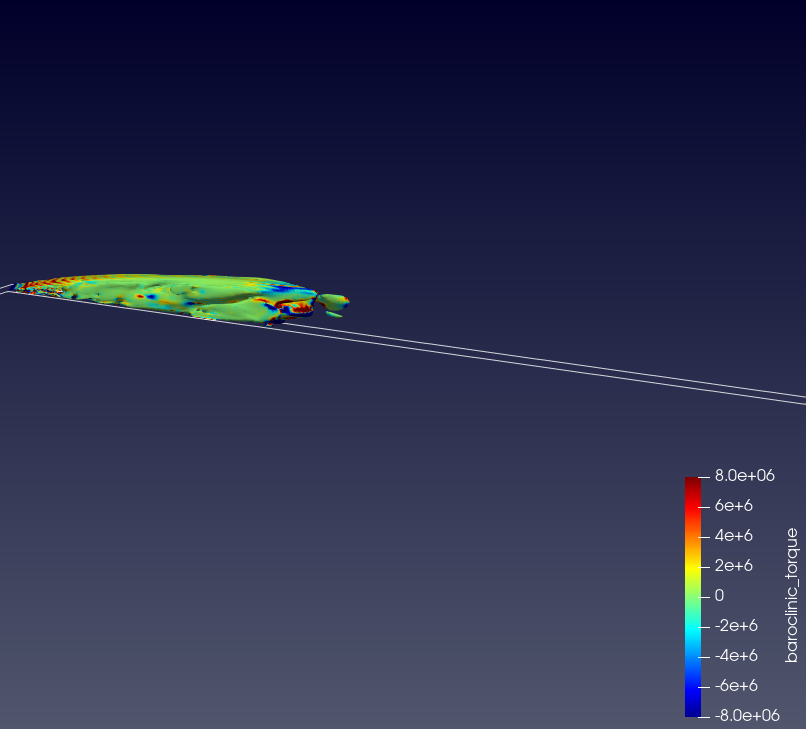}
         \label{fig:bt2}
     \end{subfigure}
        \caption{Starting top left, in clockwise direction,contours of void fraction (contour of $\alpha$ = 0.6, vortex stretching, baroclinic torque and vortex dilatation terms as the cavity reaches its maximum size.}        
        \label{fig:c2}
\end{figure}

\begin{figure*}[htbp]
\includegraphics[width=\textwidth, height=8cm]{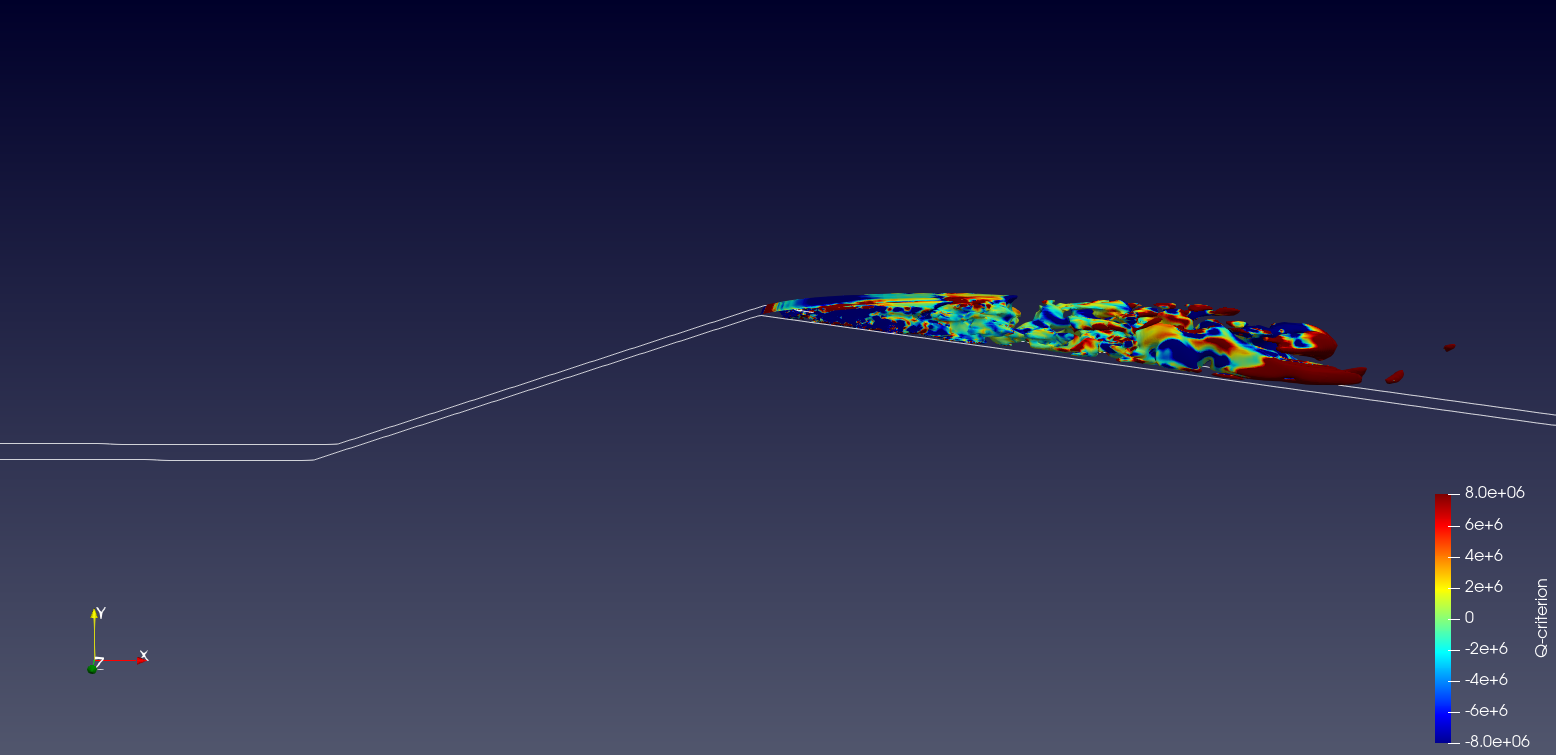}
\caption{Q-criterion contour as the cavity detaches into a secondary detached cavity}
\label{fig:cdq}
\end{figure*}

\begin{figure}[htbp]
     \centering
     \begin{subfigure}
         \centering
         \includegraphics[scale=0.35]{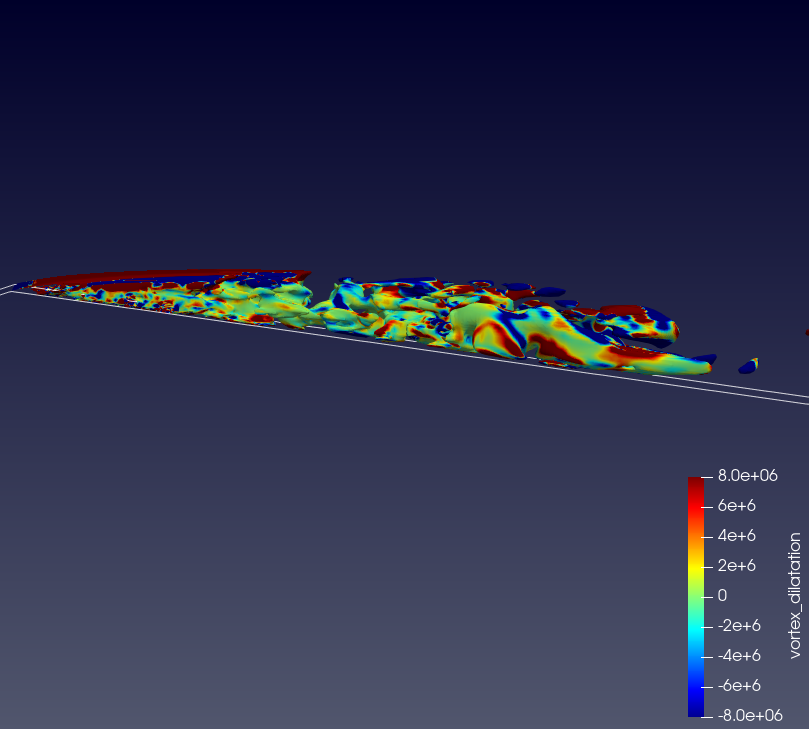}
         \label{fig:vd3}
     \end{subfigure}
     \hfill
     \begin{subfigure}
         \centering
         \includegraphics[scale=0.35]{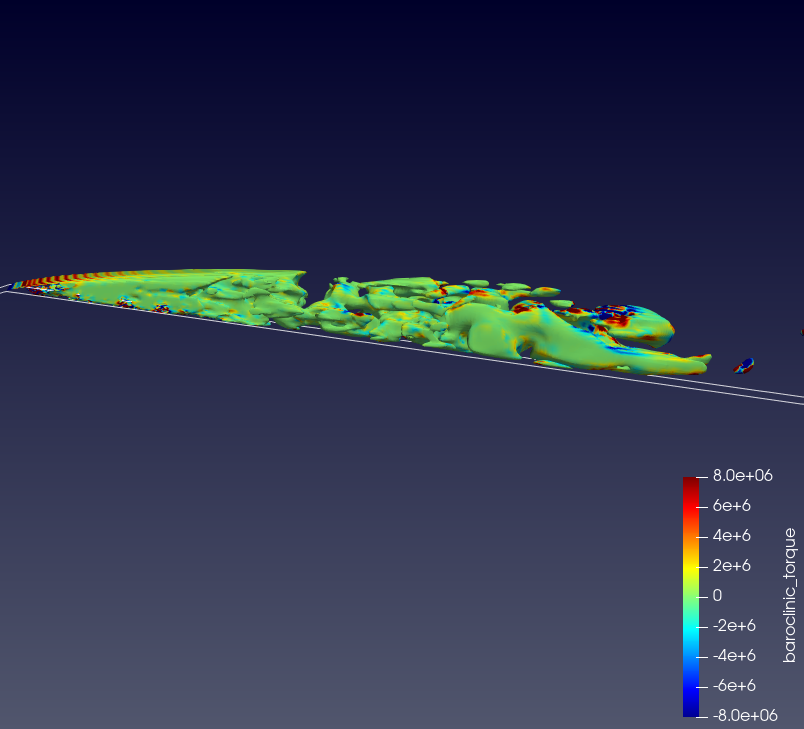}
         \label{fig:bt3}
     \end{subfigure}
        \caption{Starting top left, in clockwise direction,contours of void fraction (contour of $\alpha$ = 0.6, vortex stretching, baroclinic torque and vortex dilatation terms as the cavity detaches into a secondary detached cavity and a smaller primary cavity at the throat.}        \label{fig:cd2}
\end{figure}

Fig \ref{fig:c2} shows the cavity reaching its maximum size forming its characteristic shape. Here, the vortex stretching and dilatation terms continue to exhibit high values but are zero near the end of the cavity near the wall, where the baroclinic torque starts to show some influence. The torque also increases slightly towards the throat of the venturi. Fig \ref{fig:cd2} shows the cavity detachment process. Here, the primary cavity is broken due to a re-entrant jet rushing upwards resulting into a secondary cavity downstream and a smaller cavity at the throat. Fig \ref{fig:cdq} shows as the attached cavity breaks, the vorticity decreases with the strain rate term dominating the process. Here, the baroclinic torque starts contributing substantially to the vortex dynamics, especially at the cavity-water interface. As shown previously, the velocity experiences a 'jump' at the cavity interface and the resulting misalignment with the pressure gradients results in baroclinic torque.On the other hand, the vortex stretching and dilatation terms are mostly zero in the secondary cavity but continue to exhibit high values in the smaller primary cavity.  As the fluid element reached its maximum size in the previous stage, the vortex dilatation term reached its maximum values as well.  
\begin{figure*}[htbp]
\includegraphics[width=\textwidth, height=8cm]{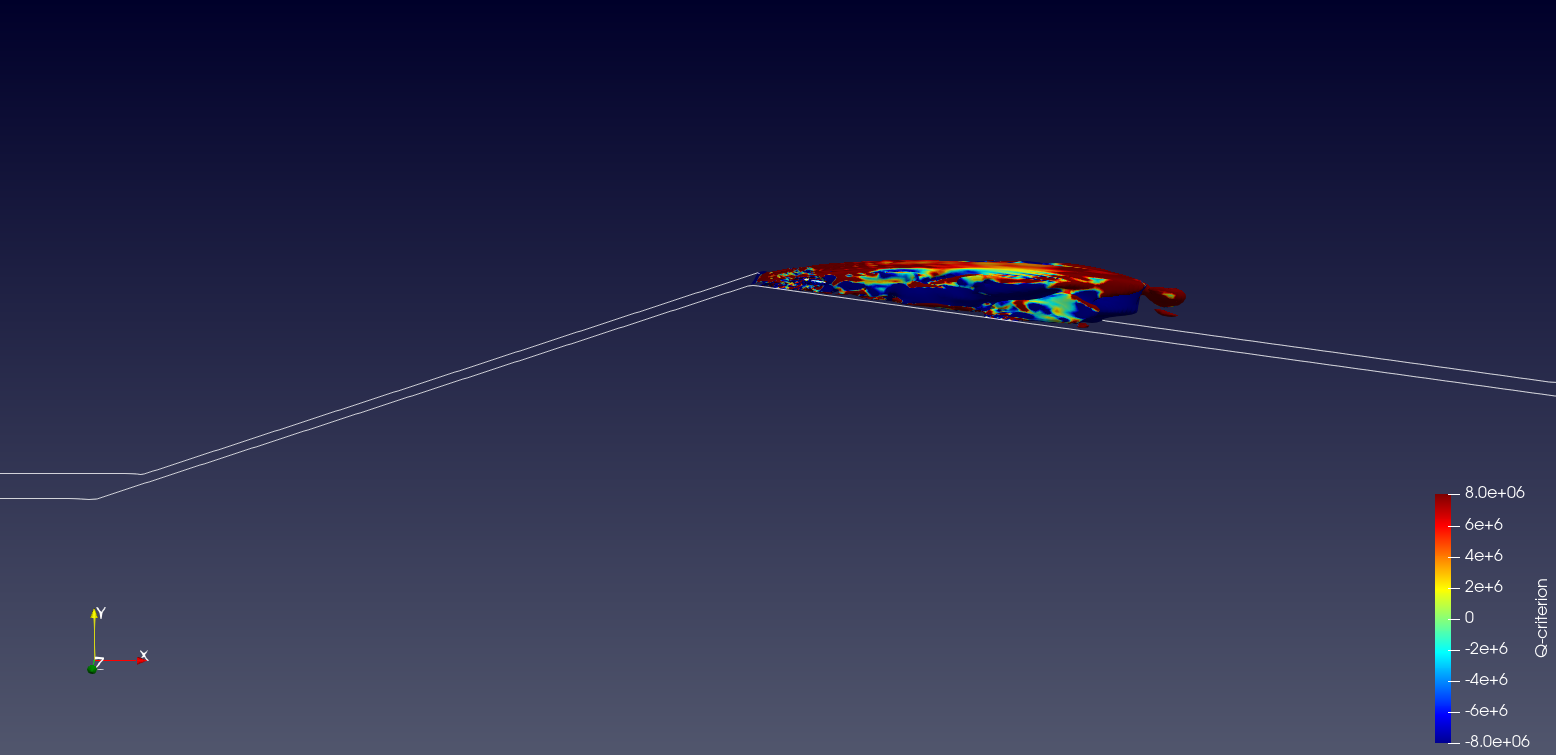}
\caption{Q-criterion contour as cavity reaches its maximum size}
\label{fig:cq}
\end{figure*}
\begin{figure}[htbp]
     \centering
     \begin{subfigure}
         \centering
         \includegraphics[scale=0.35]{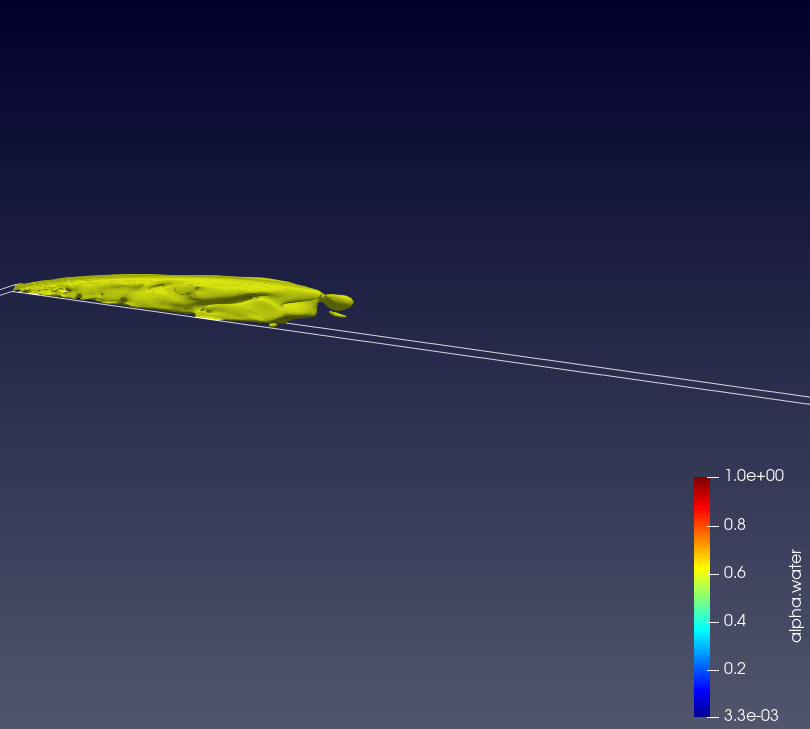}
         \label{fig:void_fraction2}
     \end{subfigure}
     \hfill
     \begin{subfigure}
         \centering
         \includegraphics[scale=0.35]{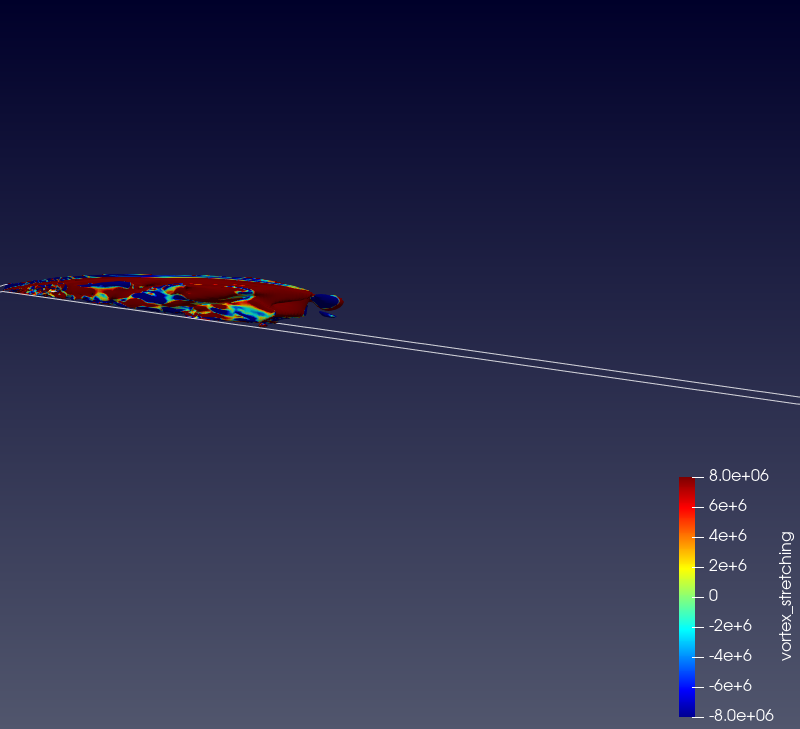}
         \label{fig:vs2}
     \end{subfigure}
\label{fig:c1}
\end{figure}

\begin{figure}[htbp]
     \centering
     \begin{subfigure}
         \centering
         \includegraphics[scale=0.35]{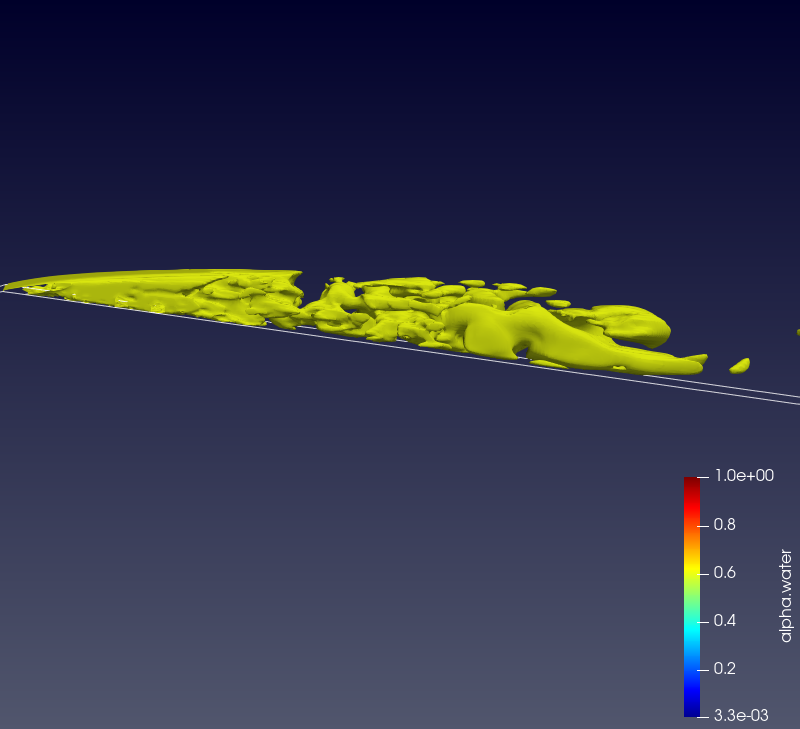}
         \label{fig:void_fraction3}
     \end{subfigure}
     \hfill
     \begin{subfigure}
         \centering
         \includegraphics[scale=0.35]{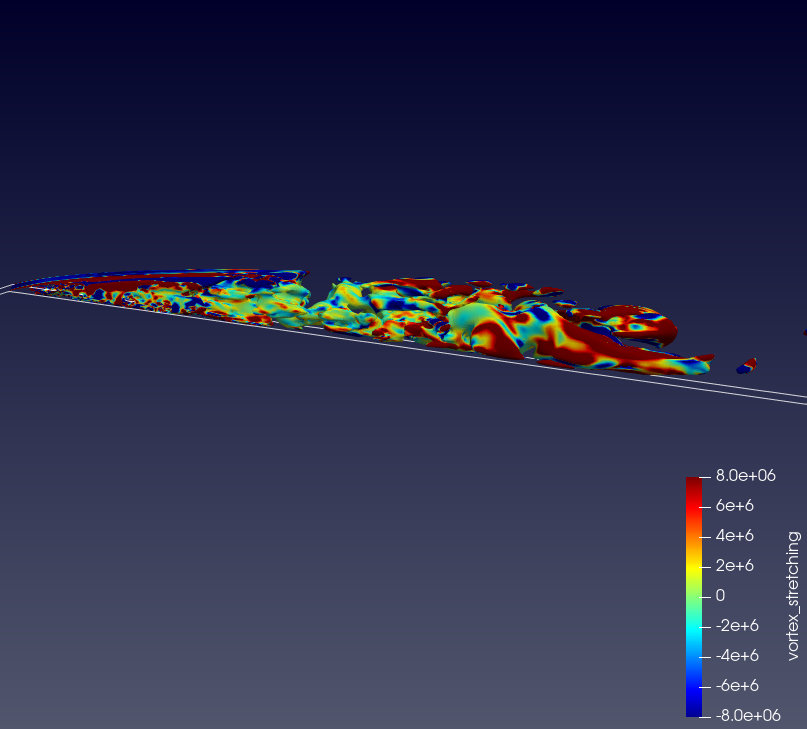}
         \label{fig:vs3}
     \end{subfigure}
\label{fig:cd1}
\end{figure}
\begin{figure*}[htbp]
\includegraphics[width=\textwidth, height=8cm]{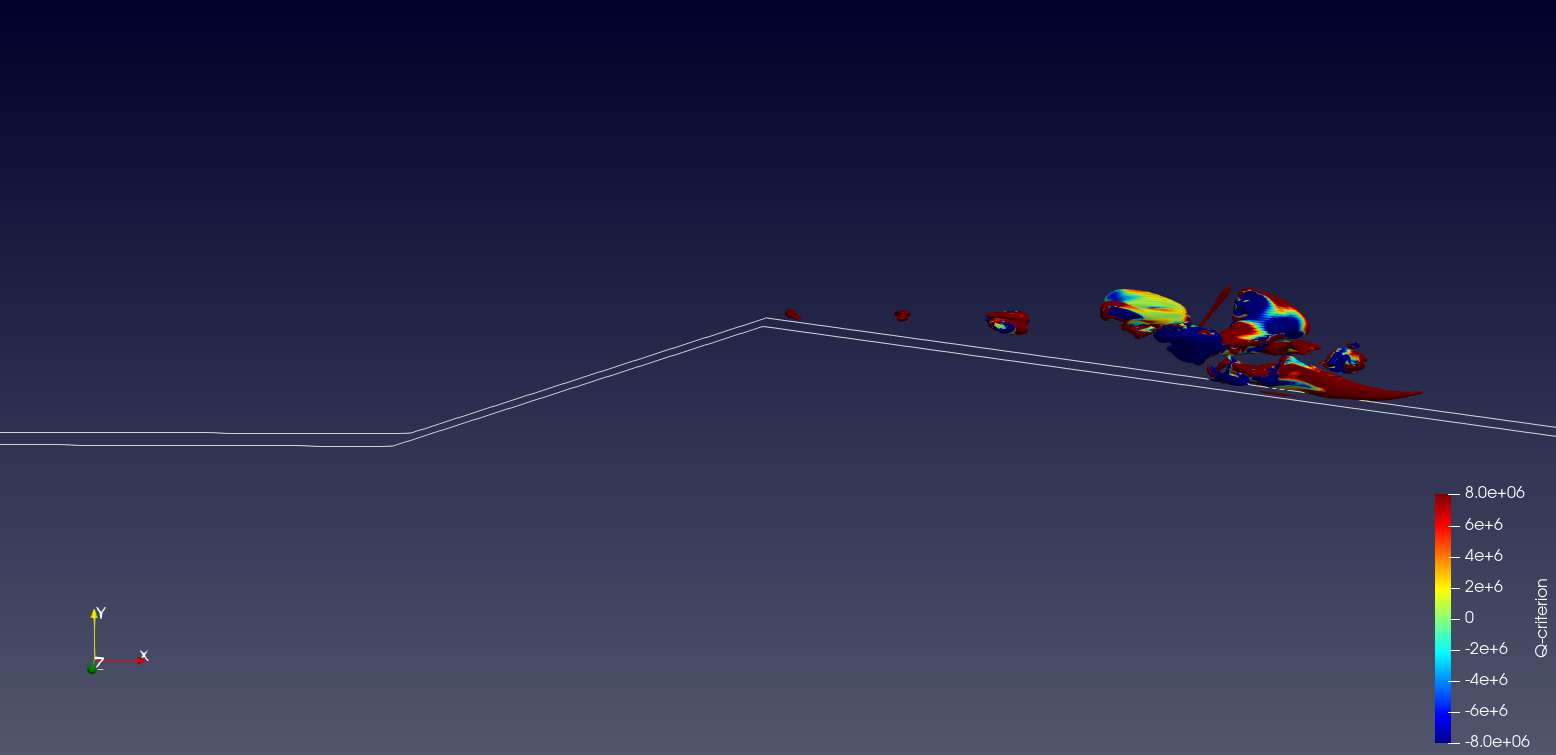}
\caption{Q-criterion contour as the secondary cavity rolls downstream}
\label{fig:nq}
\end{figure*}

\begin{figure}[htbp]
     \centering
     \begin{subfigure}
         \centering
         \includegraphics[scale=0.35]{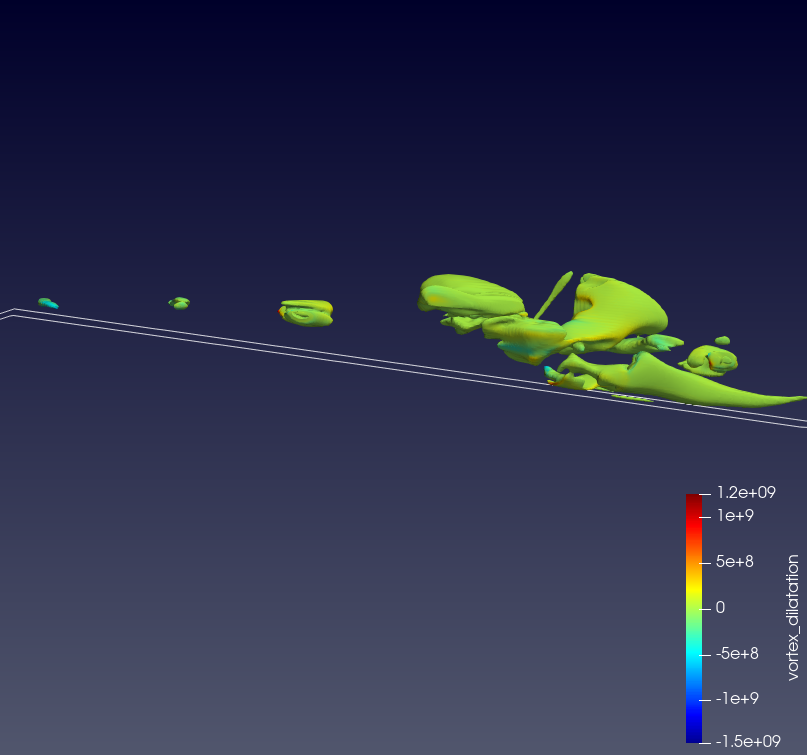}
         \label{fig:vd4}
     \end{subfigure}
     \hfill
     \begin{subfigure}
         \centering
         \includegraphics[scale=0.35]{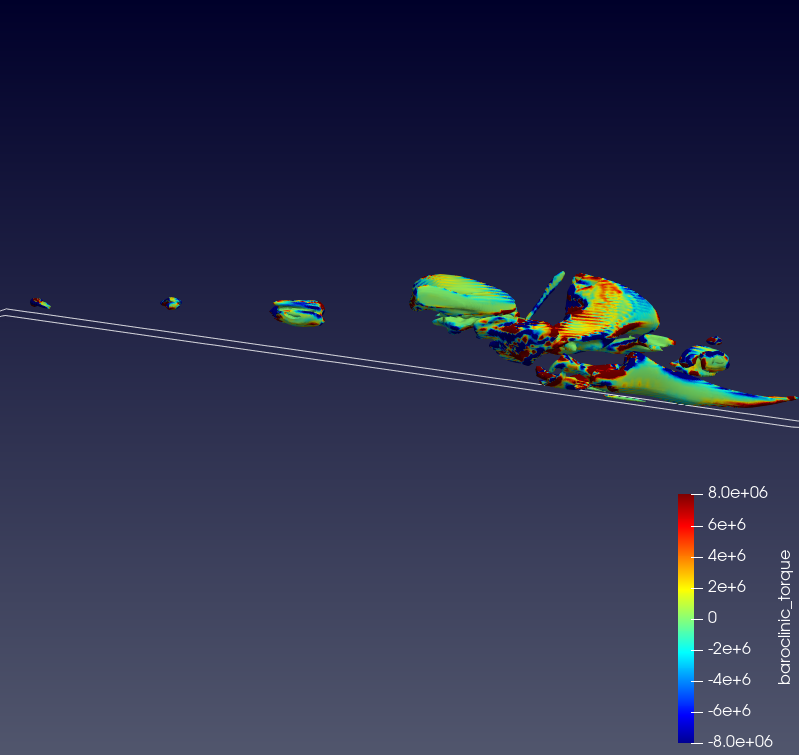}
         \label{fig:bt4}
     \end{subfigure}
        \caption{Starting top left, in clockwise direction,contours of void fraction (contour of $\alpha$ = 0.6, vortex stretching, baroclinic torque and vortex dilatation terms as the primary cavity collapses while the secondary cavity rolls up downstream, setting the stage for the following cavitating cycle}        \label{fig:n2}
\end{figure}

Fig \ref{fig:n2} shows the next stage. Here, the detached secondary cavity continues to roll-up downstream and collapse, upon exiting the low-pressure region while Fig \ref{fig:nq} shows more vorticity downstream. Here, the vortex stretching and vortex dilatation terms appear to be zero almost throughout the domain while the baroclinic torque is observed to be the consequential term for the vortex dynamics process. This corroborates the experimental findings of Laberteaux and Ceccio \citep{laberteaux2001partial}: the adverse pressure gradients formed near the cavity wake contribute significantly to the influence of baroclinic torque at the cavity collapse and shedding stage. The secondary cavity downstream was also observed in previously in Fig \ref{fig:ci2} with the growth of incipient cavity at the throat, thus starting another cavitation cycle.
In summary, the vorticity transport equation analysis shows that while there is a strong cavitation-vortex interaction, different components contribute to the vortex formation at different stages of the cavitating flow.

\begin{figure}[htbp]
     \centering
     \begin{subfigure}
         \centering
         \includegraphics[scale=0.35]{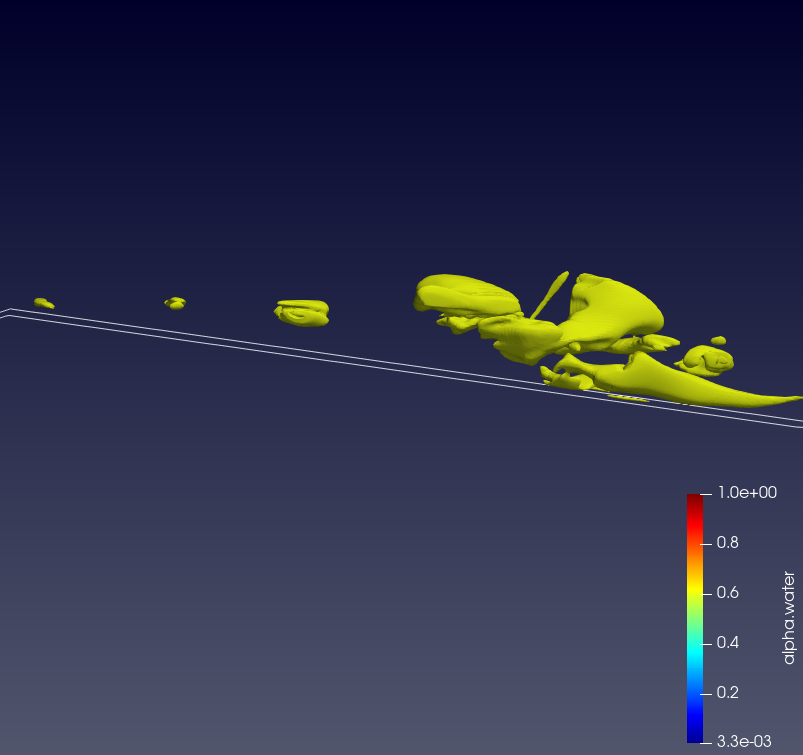}
         \label{fig:void_fraction4}
     \end{subfigure}
     \hfill
     \begin{subfigure}
         \centering
         \includegraphics[scale=0.35]{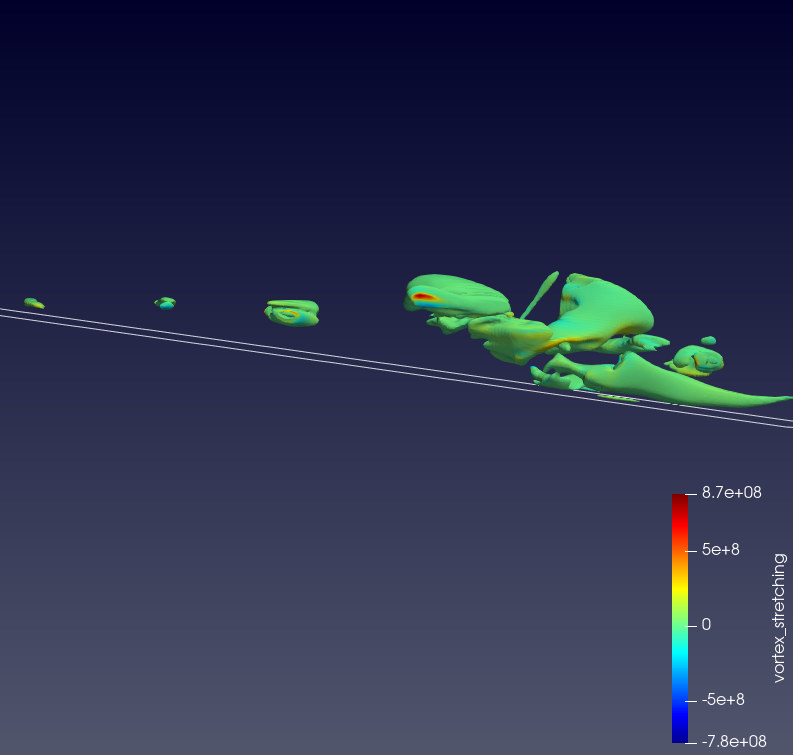}
         \label{fig:vs4}
     \end{subfigure}
\label{fig:n1}
\end{figure}

\section{Conclusions}
In this paper, the periodic cavitating flow inside a venturi-type HCR was simulated using the k-$\omega$ SST and the DES models for both 2D and 3D calculations. Regions of the HCR where cavitation manifests were refined to ensure the regions were modelled by LES rather than RANS for the DES calculations. A comprehensive analysis was conducted by comparing the simulation data with experimental data, especially on the local scale. The main conclusions are as follows:
\begin{itemize}
    \item Refining the 2D grids to ensure the cavitating flow was modelled by DES aided in the modelling of the periodic vapor shedding. However, the URANS calculations displayed unorthodox results where a primary cavity persisted at the throat throughout the entirety of the simulation. The arrow-head shape of the mean primary cavity also differed from the traditional primary cloud cavity observed in experiments and the DES simulations. The unconventional shape is a result of the re-entrant jet breaking the primary cavity at a distance much farther from the bottom wall as compared to the DES simulations. 
    \item Comparisons with local velocity profile data concluded with the models well predicting the experimental streamwise velocity data but predicting the wall direction velocity with considerable discrepancies. These discrepancies influenced further the Reynolds stress and Turbulent Kinetic Energy (TKE) profiles. While the simulations were able to predict the dynamics near the throat, significant discrepancies were observed in the downstream profiles. 
    \item The discussion was extended to 3D calculations to investigate the 3D effects of the cavitation-turbulence coupling. Local profile plots indicated that the 3D profiles were able to well-predict the turbulence statistics profiles throughout the cavity region as observed in experiments. The improvements proposed to the original DES model in the form of DDES and IDDES models visibly affected the profiles as these modifications aided in predicting TKE magnitudes identical to experimental data. The URANS simulation exhibited considerable discrepancies and under-predicted the turbulent dynamics in the profiles.  
    \item Vorticity-budget analysis on the cavitation-vortex interaction showed that the vortex stretching and vortex dilatation terms influence the vortex dynamics significantly in a cavitating shedding cycle. However, the shedding of the secondary cavity and its subsequent collapse was dominated by the baroclinic torque due to the resulting adverse pressure gradients.
    \item While the 3D simulations were able to well predict the flow dynamics associated with unsteady cavitation, their high computing costs pose a significant challenge. We propose employing frameworks where efficient URANS turbulence models are augmented with techniques driven by experimental data for greater accuracy and reducing or even retaining the computational cost. 
\end{itemize}
\section{Funding}
This work was supported by the Office of Naval Research, USA [grant number N00014-18-S-B001]. The authors would like to thank the ONR proposal manager Dr. Ki-Han Kim for his support.

\section{Data Availability}

The data that support the findings of this study are available from the corresponding author upon reasonable request.

\bibliographystyle{elsarticle-num} 
\bibliography{references}

\begin{thebibliography}{10}
\expandafter\ifx\csname url\endcsname\relax
  \def\url#1{\texttt{#1}}\fi
\expandafter\ifx\csname urlprefix\endcsname\relax\def\urlprefix{URL }\fi
\expandafter\ifx\csname href\endcsname\relax
  \def\href#1#2{#2} \def\path#1{#1}\fi

\bibitem{hutli2017experimental}
E.~Hutli, M.~S. Nedeljkovic, A.~Bony{\'a}r, D.~L{\'e}gr{\'a}dy, Experimental study on the influence of geometrical parameters on the cavitation erosion characteristics of high speed submerged jets, Experimental Thermal and Fluid Science 80 (2017) 281--292.

\bibitem{chuah2017kinetic}
L.~F. Chuah, J.~J. Kleme{\v{s}}, S.~Yusup, A.~Bokhari, M.~M. Akbar, Z.~K. Chong, Kinetic studies on waste cooking oil into biodiesel via hydrodynamic cavitation, Journal of Cleaner Production 146 (2017) 47--56.

\bibitem{laosuttiwong2018performance}
T.~Laosuttiwong, K.~Ngaosuwan, W.~Kiatkittipong, D.~Wongsawaeng, P.~Kim-Lohsoontorn, S.~Assabumrungrat, Performance comparison of different cavitation reactors for biodiesel production via transesterification of palm oil, Journal of cleaner production 205 (2018) 1094--1101.

\bibitem{wang2021hydrodynamic}
B.~Wang, H.~Su, B.~Zhang, Hydrodynamic cavitation as a promising route for wastewater treatment--a review, Chemical Engineering Journal 412 (2021) 128685.

\bibitem{saxena2018enhanced}
S.~Saxena, V.~K. Saharan, S.~George, Enhanced synergistic degradation efficiency using hybrid hydrodynamic cavitation for treatment of tannery waste effluent, Journal of Cleaner Production 198 (2018) 1406--1421.

\bibitem{sun2021multi}
X.~Sun, Z.~Yang, X.~Wei, Y.~Tao, G.~Boczkaj, J.~Y. Yoon, X.~Xuan, S.~Chen, Multi-objective optimization of the cavitation generation unit structure of an advanced rotational hydrodynamic cavitation reactor, Ultrasonics Sonochemistry 80 (2021) 105771.

\bibitem{xia2024numerical}
G.~Xia, W.~You, S.~Manickam, J.~Y. Yoon, X.~Xuan, X.~Sun, Numerical simulation of cavitation-vortex interaction mechanism in an advanced rotational hydrodynamic cavitation reactor, Ultrasonics Sonochemistry (2024) 106849.

\bibitem{mittal2023intensifying}
R.~Mittal, V.~V. Ranade, Intensifying extraction of biomolecules from macroalgae using vortex based hydrodynamic cavitation device, Ultrasonics Sonochemistry 94 (2023) 106347.

\bibitem{sarvothaman2024evaluating}
V.~P. Sarvothaman, S.~R. Kulkarni, J.~Subburaj, S.~L. Hariharan, V.~K. Velisoju, P.~Casta{\~n}o, P.~Guida, D.~M. Prabhudharwadkar, W.~L. Roberts, Evaluating performance of vortex-diode based hydrodynamic cavitation device scale and pressure drop using coumarin dosimetry, Chemical Engineering Journal 481 (2024) 148593.

\bibitem{ge2022dynamic}
M.~Ge, P.~Manikkam, J.~Ghossein, R.~K. Subramanian, O.~Coutier-Delgosha, G.~Zhang, Dynamic mode decomposition to classify cavitating flow regimes induced by thermodynamic effects, Energy 254 (2022) 124426.

\bibitem{mukherjee2022design}
A.~Mukherjee, S.~Chalicheemala, S.~Roy, A.~Mullick, S.~De, A.~Roy, S.~Moulik, Design, development and performance evaluation of skid-mounted pilot wastewater treatment and resource recovery unit for mechanical scavenging vehicle, Journal of Cleaner Production 371 (2022) 133564.

\bibitem{callenaere2001cavitation}
M.~Callenaere, J.-P. Franc, J.-M. Michel, M.~Riondet, The cavitation instability induced by the development of a re-entrant jet, Journal of Fluid Mechanics 444 (2001) 223--256.

\bibitem{lyu2021modeling}
X.~Lyu, Y.~Zhu, C.~Zhang, X.~Hu, N.~A. Adams, Modeling of cavitation bubble cloud with discrete lagrangian tracking, Water 13~(19) (2021) 2684.

\bibitem{wang2022numerical}
Z.~Wang, H.~Cheng, B.~Ji, Numerical prediction of cavitation erosion risk in an axisymmetric nozzle using a multi-scale approach, Physics of Fluids 34~(6) (2022).

\bibitem{merkle1998computational}
C.~L. Merkle, Computational modelling of the dynamics of sheet cavitation, in: Proc. of the 3rd Int. Symp. on Cavitation, Grenoble, France, 1998, 1998.

\bibitem{kunz2000preconditioned}
R.~F. Kunz, D.~A. Boger, D.~R. Stinebring, T.~S. Chyczewski, J.~W. Lindau, H.~J. Gibeling, S.~Venkateswaran, T.~Govindan, A preconditioned navier--stokes method for two-phase flows with application to cavitation prediction, Computers \& Fluids 29~(8) (2000) 849--875.

\bibitem{schnerr2001physical}
G.~H. Schnerr, J.~Sauer, Physical and numerical modeling of unsteady cavitation dynamics, in: Fourth international conference on multiphase flow, Vol.~1, ICMF New Orleans New Orleans, LO, USA, 2001.

\bibitem{gnanaskandan2016large}
A.~Gnanaskandan, K.~Mahesh, Large eddy simulation of the transition from sheet to cloud cavitation over a wedge, International Journal of Multiphase Flow 83 (2016) 86--102.

\bibitem{trummler2020investigation}
T.~Trummler, S.~J. Schmidt, N.~A. Adams, Investigation of condensation shocks and re-entrant jet dynamics in a cavitating nozzle flow by large-eddy simulation, International Journal of Multiphase Flow 125 (2020) 103215.

\bibitem{coutier2003evaluation}
O.~Coutier-Delgosha, R.~Fortes-Patella, J.-L. Reboud, Evaluation of the turbulence model influence on the numerical simulations of unsteady cavitation, J. Fluids Eng. 125~(1) (2003) 38--45.

\bibitem{vaca2023numerical}
D.~Vaca-Revelo, A.~Gnanaskandan, Numerical assessment of the condensation shock mechanism in sheet to cloud cavitation transition, International Journal of Multiphase Flow 169 (2023) 104616.

\bibitem{spalart1997comments}
P.~R. Spalart, Comments on the feasibility of les for wings and on the hybrid rans/les approach, in: Proceedings of the First AFOSR International Conference on DNS/LES, 1997, 1997, pp. 137--147.

\bibitem{spalart2006new}
P.~R. Spalart, S.~Deck, M.~L. Shur, K.~D. Squires, M.~K. Strelets, A.~Travin, A new version of detached-eddy simulation, resistant to ambiguous grid densities, Theoretical and computational fluid dynamics 20 (2006) 181--195.

\bibitem{shur2008hybrid}
M.~L. Shur, P.~R. Spalart, M.~K. Strelets, A.~K. Travin, A hybrid rans-les approach with delayed-des and wall-modelled les capabilities, International journal of heat and fluid flow 29~(6) (2008) 1638--1649.

\bibitem{bensow2011simulation}
R.~E. Bensow, Simulation of the unsteady cavitation on the delft twist11 foil using rans, des and les, in: Second international symposium on marine propulsors, 2011.

\bibitem{sedlar2016numerical}
M.~Sedlar, B.~Ji, T.~Kratky, T.~Rebok, R.~Huzl{\'\i}k, Numerical and experimental investigation of three-dimensional cavitating flow around the straight naca2412 hydrofoil, Ocean Engineering 123 (2016) 357--382.

\bibitem{pipp2021challenges}
P.~Pipp, M.~Ho{\v{c}}evar, M.~Dular, Challenges of numerical simulations of cavitation reactors for water treatment--an example of flow simulation inside a cavitating microchannel, Ultrasonics sonochemistry 77 (2021) 105663.

\bibitem{apte2023numerical}
D.~Apte, M.~Ge, O.~Coutier-Delgosha, Numerical investigation of a cavitating nozzle for jetting and rock erosion based on different turbulence models, Geoenergy Science and Engineering 231 (2023) 212300.

\bibitem{simpson2019modeling}
A.~Simpson, V.~V. Ranade, Modeling hydrodynamic cavitation in venturi: Influence of venturi configuration on inception and extent of cavitation, AIChE Journal 65~(1) (2019) 421--433.

\bibitem{bimestre2020theoretical}
T.~A. Bimestre, J.~A.~M. J{\'u}nior, C.~A. Botura, E.~Canettieri, C.~E. Tuna, Theoretical modeling and experimental validation of hydrodynamic cavitation reactor with a venturi tube for sugarcane bagasse pretreatment, Bioresource technology 311 (2020) 123540.

\bibitem{fang2020numerical}
L.~Fang, W.~Li, Q.~Li, Z.~Wang, Numerical investigation of the cavity shedding mechanism in a venturi reactor, International Journal of Heat and Mass Transfer 156 (2020) 119835.

\bibitem{reboud1998two}
J.-L. Reboud, B.~Stutz, O.~Coutier, Two phase flow structure of cavitation: experiment and modeling of unsteady effects, in: 3rd international symposium on cavitation CAV1998, Grenoble, France, Vol.~26, 1998, pp. 1--8.

\bibitem{dutta2021novel}
N.~Dutta, P.~Kopparthi, A.~K. Mukherjee, N.~Nirmalkar, G.~Boczkaj, Novel strategies to enhance hydrodynamic cavitation in a circular venturi using rans numerical simulations, Water research 204 (2021) 117559.

\bibitem{malekshah2022dissolved}
E.~H. Malekshah, W.~Wr{\'o}blewski, M.~Majkut, Dissolved air effects on three-phase hydrodynamic cavitation in large scale venturi-experimental/numerical analysis, Ultrasonics Sonochemistry 90 (2022) 106199.

\bibitem{liu2024design}
H.~Liu, C.~Li, S.~Zhao, H.~Zhu, Y.~Huang, W.~He, Y.~Zhao, Y.~Li, K.~Guo, Design and structural parameter optimization of venturi-type microbubble reactor for wastewater treatment by cfd simulation, Journal of Flow Chemistry (2024) 1--16.

\bibitem{guo2024numerical}
M.~Guo, Y.~Lu, H.~Xue, P.~L. Show, J.~Y. Yoon, X.~Sun, Numerical simulation on the enhancement mechanism of the distinctive division baffle on the performance of a venturi for process intensification, Journal of Water Process Engineering 57 (2024) 104719.

\bibitem{spalart1992one}
P.~Spalart, S.~Allmaras, A one-equation turbulence model for aerodynamic flows, in: 30th aerospace sciences meeting and exhibit, 1992, p. 439.

\bibitem{travin2002physical}
A.~Travin, M.~Shur, M.~Strelets, P.~Spalart, Physical and numerical upgrades in the detached-eddy simulation of complex turbulent flows, in: Advances in LES of Complex Flows: Proceedings of the Euromech Colloquium 412, held in Munich, Germany 4--6 October 2000, Springer, 2002, pp. 239--254.

\bibitem{menter2003ten}
F.~R. Menter, M.~Kuntz, R.~Langtry, et~al., Ten years of industrial experience with the sst turbulence model, Turbulence, heat and mass transfer 4~(1) (2003) 625--632.

\bibitem{agarkoti2023pilot}
C.~Agarkoti, S.~K. Gujar, P.~R. Gogate, A.~B. Pandit, Pilot scale degradation of sulfamerazine using different venturi based hydrodynamic cavitation and ultrasound reactors in combination with oxidation processes, Journal of Environmental Chemical Engineering 11~(3) (2023) 109857.

\bibitem{thakkar2022multi}
K.~Thakkar, S.~S. Kachhwaha, P.~Kodgire, Multi-response optimization of transesterification reaction for biodiesel production from castor oil assisted by hydrodynamic cavitation, Fuel 308 (2022) 121907.

\bibitem{dehkordi2017cfd}
P.~B. Dehkordi, L.~P.~M. Colombo, M.~Guilizzoni, G.~Sotgia, Cfd simulation with experimental validation of oil-water core-annular flows through venturi and nozzle flow meters, Journal of Petroleum science and Engineering 149 (2017) 540--552.

\bibitem{weller1998tensorial}
H.~G. Weller, G.~Tabor, H.~Jasak, C.~Fureby, A tensorial approach to computational continuum mechanics using object-oriented techniques, Computers in physics 12~(6) (1998) 620--631.

\bibitem{dittakavi2010large}
N.~Dittakavi, A.~Chunekar, S.~Frankel, Large eddy simulation of turbulent-cavitation interactions in a venturi nozzle, Journal of Fluids Engineering 132~(12) (2010).

\bibitem{laberteaux2001partial}
K.~Laberteaux, S.~Ceccio, Partial cavity flows. part 1. cavities forming on models without spanwise variation, Journal of Fluid Mechanics 431 (2001) 1--41.

\end{thebibliography}
\end{document}